\documentclass[useAMS,usenatbib]{mn2e}

\usepackage{epsf,rotating}
\usepackage{amsmath}
\usepackage{subfigure}

\newcommand{\kmsmpc}{\kms\;{\rm Mpc}^{-1}}

\newcommand{\hkpc}{h^{-1}{\rm kpc}}
\newcommand{\hmpc}{h^{-1}{\rm Mpc}}

\newcommand{\kms}{\;{\rm km}\,{\rm s}^{-1}}

\newcommand{\cmc}{\;{\rm cm}^{-3}}

\newcommand{\msolar}{\;{\rm M}_{\odot}}

\newcommand{\gad}{{\sc Gadget-2}}

\newcommand{\ion}[2]{\hbox{#1\,{\sc #2}}}
\newcommand{\vw}{v_{\rm w}}

\newcommand{\fgas}{f_{\rm gas}}

\title[Galaxy Evolution in Simulations with Outflows I: $M_*$ and SFR]{Galaxy Evolution in Cosmological Simulations With Outflows I: Stellar Masses and Star Formation Rates}

\author[Dav\'e, Oppenheimer, Finlator]{
\parbox[t]{\textwidth}{\vspace{-1cm}
Romeel Dav\'e$^1$, Benjamin D. Oppenheimer$^2$, Kristian Finlator$^3$}
\\\\$^1$ Astronomy Department, University of Arizona, Tucson, AZ 85721, USA
\\$^2$ Veni Fellow; Leiden Observatory, Leiden University, PO Box 9513, 2300 RA Leiden, Netherlands
\\$^3$ Hubble Fellow; Physics Department, University of California, Santa Barbara, CA 93106, USA
}

\begin{document}

\maketitle

 \begin{abstract}
We examine the growth of the stellar content of galaxies
from $z=3\rightarrow 0$ in cosmological hydrodynamic simulations
incorporating parameterised galactic outflows.  Without outflows,
galaxies overproduce stellar masses ($M_*$) and star formation rates
(SFRs) compared to observations.  Winds introduce a three-tier form for
the galaxy stellar mass and star formation rate functions, where the
middle tier depends on differential (i.e. mass-dependent) recycling of
ejected wind material back into galaxies.  A tight $M_*-$SFR relation
is a generic outcome of all these simulations, and its evolution is
well-described as being powered by cold accretion, although current
observations at $z\ga 2$ suggest that star formation in small early
galaxies must be highly suppressed.  Roughly one-third of $z=0$
galaxies at masses below $M^\star$ are satellites, and star formation
in satellites is not much burstier than in centrals.  All models fail
to suppress star formation and stellar mass growth in massive galaxies
at $z\la 2$, indicating the need for an external quenching mechanism
such as black hole feedback.  All models also fail to produce dwarfs as
young and rapidly star-forming as observed.  An outflow model following
scalings expected for momentum-driven winds broadly matches observed
galaxy evolution around $M^\star$ from $z=0-3$, which is a significant
success since these galaxies dominate cosmic star formation, but the
failures at higher and lower masses highlight the challenges still faced
by this class of models.  We argue that central star-forming galaxies
are well-described as living in a slowly-evolving equilibrium between
inflows from gravity and recycled winds, star formation, and strong
and ubiquitous outflows that regulate how much inflow forms into stars.
Star-forming galaxy evolution is thus primarily governed by the continual
cycling of baryons between galaxies and intergalactic gas.
\end{abstract}


\section{Introduction} 

Galaxies are the primary tracers by which astronomers map and measure
the Universe, so understanding how galaxies form and evolve is a
longstanding central goal of astronomy.  The physics of galaxy
formation involves a diverse set of processes spanning an enormous
dynamic range, from black holes and stellar physics on sub-parsec
scales to cosmology and large-scale structure on gigaparsec scales.
As a consequence, despite rapidly advancing observations of galaxies
across cosmic time, a comprehensive theory for the formation and
evolution of galaxies that can explain all their observed properties
remains elusive.

As galaxies are essentially collections of stars, understanding the
growth of the stellar component of galaxies is of fundamental
importance.  Recent observations have probed the star formation
rates and stellar masses of galaxies out to redshift $z\sim 7$.
Key trends have emerged.  For instance, fairly massive galaxies are
in place at early epochs, and galaxies exhibit ``downsizing" in the
sense that more massive galaxies form their stars earlier and have
lower birthrates today.  While some claim that these trends are in
conflict with expectations from hierarchical structure formation,
they are actually not.  Since the most massive galaxies form within
the largest mass density peaks, they collapse first and start forming
stars earlier than lower-mass systems~\citep{dav06b,nei06}.  The
collapse is driven by gravitational instability and strong radiative
cooling that (left unchecked) grow galaxies very rapidly at early
epochs, so that models without strong feedback yield galaxies that
are {\it too large} at high redshifts~\citep[e.g.][]{dav06}.  It
is important to note that no confirmed high-$z$ galaxy yet observed
is too large to be produced in a hierarchical model, and in fact
strong feedback appears to be necessary to suppress galaxy formation
from the earliest epochs.  The evolution of galaxy stellar masses
and star formation rates therefore provide valuable insights into
the processes, particularly related to feedback, that drive galaxy
formation from the highest redshift until today.

The canonical scenario for galaxy formation posits that galaxies
assemble first as rotationally-supported disks cooling out of
virial-temperature gas within a dark matter
halo~\citep[e.g.][]{ree77,whi78,whi91}, and then such objects merge
to form larger and earlier-type systems~\citep[e.g.][]{mih96}.
Naively, one expects then that star-forming disks would be surrounded
by hot gaseous halos~\citep{ben00}, and that early low-mass objects
would tend to be disky while more massive objects would be ellipticals.
Recent observations, however, do not clearly support this paradigm.
For instance, star-forming disks and hot (X-ray emitting) gaseous
halos are rarely if ever found together~\citep[e.g.][though see
\citealt{cra10} and references therein]{ras09}.  Observed dynamics
of $z\sim 1-2$ galaxies indicate that larger objects tend to be
disky, while smaller ones are dispersion-dominated~\citep{for09}.
Furthermore, galaxies appear to be consuming their gas rapidly
enough that fuel must be constantly replenished~\citep{tac10,pap10},
but merger-induced starbursts do not appears to be driving global
star formation~\citep{bri04,noe07,jog09}.  Hence there is some doubt
that the canonical halo/merger-centric view provides the appropriate
framework for galaxy formation, despite the wide-ranging successes
that models based on this have enjoyed~\citep[e.g.][]{ben10}.

Improvements in cosmological hydrodynamic simulations over the past
decade have made them a competitive approach towards understanding
the physics of galaxy formation in a hierarchical structure formation
context.  The complex three-dimensional geometry of large-scale
structure is self-consistently evolved together with the baryonic
components that make up observable galaxies and the co-evolving
intergalactic medium (IGM).  These simulations have long suggested
that accretion onto galaxies is powered by cold, filamentary streams
that connect to the cosmic web on larger scales~\citep{ker05,dek09},
and that most accreted gas fueling star formation never heats to
the virial temperature~\citep{kat91,dek06}.  Gas replenishment from
the IGM is a natural outcome of this model, and generally occurs
smoothly rather than being driven by major
mergers~\citep[e.g.][]{mur02,ker05,vdvoort10}.  These simulations
provide many insights into how galaxies form and grow, some of which
are in common with the traditional scenario, while some differ.

Recently, simulations incorporating strong and ubiquitous outflows
have had success matching a wide range of data on
galaxies~\citep[e.g.][]{dav06,fin08,opp10} and the
IGM~\citep[e.g.][]{opp06,opp08,opp09,opp09b}.  The outflow models
in these simulations were originally tuned to enrich the IGM in
accord with observations of $z\sim 2-4$ \ion{C}{iv}
absorbers~\citep{opp06}.  The favoured scalings turn out to be those
expected for momentum-conserved winds~\citep{mur05,zha10}, namely
that the wind speed scales with circular (or escape) velocity, while
the amount of mass ejected per unit star formation (i.e. the mass
loading factor) scales inversely with it.  This provides sufficiently
early IGM enrichment by expelling the majority of metals formed in
small high-redshift galaxies, while having moderate wind speeds
that does not to over-heat the IGM~\citep{opp06}.  In subsequent work
we showed that an outflow model with similar scalings consistently
does as well or better than any other outflow model that we have
tried at matching a wide range of IGM properties.  We note that the
actual implementation of this wind model in our simulations has
evolved owing to advances in modeling techniques and changes in
cosmology~\citep{opp08,opp10}.  Successes include very early ($z\ga
5$) enrichment~\citep{opp09b}, \ion{O}{vi} absorbers at low-$z$~\citep[with
the introduction of a physically-motivated IGM turbulence
model][]{opp09}, the metallicities in intragroup and intracluster
gas~\citep{dav08b}, and the overall cosmic metal
distribution~\citep[including the so-called ``missing metals
problem";][]{dav07}.

Examining galaxy properties, we again found that an outflow model
following similar scalings was the most successful amongst the ones
we tried.  Successes included matching the mass-metallicity relation
at $z=2$~\citep{fin08}, the shape of the present-day stellar mass
function~\citep{opp10} below $M^\star$, damped Ly$\alpha$ absorber
kinematics~\citep{hon10}, and observations of very high-$z$ ($z\ga
4$) galaxies~\citep{fin06,dav06,bou07,fin10}.  This is not to say
that this particular outflow model matches all the data; in fact,
it does not, it simply does better than others attempted, including
the oft-used \citet{spr03b} model.  Moreover, the implementation
remains highly heuristic, and may be impacted by numerical
effects~\citep[e.g. see discussion in][]{opp10}.  Nevertheless, the
consistent (comparative) success of an outflow model with momentum-driven
wind scalings suggests that it captures some essential ingredients
required for a successful model of galactic outflows, and motivates
a more detailed investigation of the implications of such outflows.

Two key predictions from models incorporating outflows with such
scalings are that (i) the total amount of material ejected from
galaxies significantly exceeds that going into stars~\citep[hereafter
OD08]{opp08}; and (ii) the ejected material often re-accretes onto
galaxies~\citep{opp10}.  The strong and ubiquitous ejection and
re-accretion of material suggest that galaxies are best viewed as
evolving within an ecosystem of surrounding intergalactic gas with
which they continually exchange mass, energy, and metals.

While the above results are encouraging, these simulations have yet
to be comprehensively tested against observations of galaxy evolution
from the present epoch back into the early universe.  Such a
comparison would provide the most stringent test to date of how
well these simulations can reproduce the observed Universe, and
would better highlight the current successes and failures of galaxy
evolution simulations.  This series of papers aims to understand
how basic galaxy statistics and scaling relations between their
stellar, gas, and metal content constrain the physics of galaxy
formation.  In Paper I (this paper) we focus on the statistics of
and scaling relations between stellar mass and star formation rate.
In Paper II (Dav\'e, Finlator, \& Oppenheimer, in prep.) we will
additionally investigate the gas and metal content of galaxies.
The primary goal of these works is to understand and characterise
the key physical processes that drive the observable properties of
galaxies across cosmic time.

In these papers, we will demonstrate that for a particular choice
of outflow model, namely outflows following scalings as expected
for momentum-driven winds, the simulations can reproduce key
observations of scaling relations between stars, gas, and metals
for large star-forming galaxies that dominate cosmic star formation.
This is a first for cosmological hydrodynamic simulations.  However,
all simulations we examine here fail to match selected galaxy
properties at high and low masses, indicating that additional or
different physics is required even for our most successful model.
Using insights gained from the simulations, we show that relations
between basic galaxy constituents can be broadly understood through
a simple scenario in which inflows are balanced by outflows and
star formation.  This contrasts with the canonical halo/merger-centric
model of galaxy formation; in our scenario, the dark matter halo's
virial radius and galaxy mergers play a sub-dominant role in the
evolution of star-forming systems.  We argue that galactic outflows
are the key moderator of the stellar content, star formation rate,
and gas and metal evolution in galaxies at all epochs.  In turn,
observations of these properties provide valuable constraints on
the cosmic ecosystem within which galaxies form and grow.

This paper is organised as follows. In \S\ref{sec:sims} we describe
our hydrodynamic simulations including our galactic outflow model.
In \S\ref{sec:msfrfcn} we investigate the galaxy stellar mass and
star formation rate functions, and explain the three-tier behavior
generically seen in wind simulations by invoking differential wind
recycling.  In \S\ref{sec:ssfr} we look at the scaling relation
between SFR and $M_*$ and its evolution to high-$z$, emphasizing
its connection with the matter accretion rate into halos.  In
\S\ref{sec:sat} we further break down these trends in terms of
centrals vs. satellite galaxies, and raise some issues regarding
observed dwarf galaxies that none of the models reproduce.  Finally,
we summarise and discuss the broader implications of our work in
\S\ref{sec:summary}.

\section{Simulations}\label{sec:sims}

\subsection{Code}

Our simulations are run with an extended version of the \gad~N-body
+ Smoothed Particle Hydrodynamic (SPH) code \citep[OD08]{spr05}.
We assume a $\Lambda$CDM cosmology using the cosmological parameters
based on recent WMAP results~\citep{hin09}: $\Omega_{\rm M}=0.28$,
$\Omega_{\rm \Lambda}=0.72$, $h\equiv H_0/(100 \kmsmpc)=0.7$, a
primordial power spectrum index $n=0.95$, an amplitude of the mass
fluctuations scaled to $\sigma_8=0.82$, and $\Omega_b=0.046$.  We
refer to these runs as the r-series, where our general naming
convention for a simulation run is r[{\it boxsize}]n[{\it
particles/side}][{\it wind model}].  Our primary runs use the boxsize
of $48\hmpc$ on a side with $384^3$ dark matter and $384^3$ gas
particles.  To expand our dynamic range we also run two additional
sets of simulations that are identical except that they have box
sizes of $24\hmpc$ and $48\hmpc$, and have only $256^3$ particles
of each species.  The particle masses and gravitational softening
lengths are listed in Table~\ref{tab:sims}.

An overview of the \gad~code can be found in \S2 of K09a.  Additions
to the public version of the code include cooling processes using
the primordial abundances as described in \citet{kat96} and metal-line
cooling as described in \citet{opp06}.  Star formation is modeled
using a subgrid recipe introduced in \citet{spr03a} where a gas
particle above a critical density is modeled as a fraction of cold
clouds embedded in a warm ionised medium following \citet{mck77}.
Star formation (SF) follows a Schmidt law \citep{sch59} with the
SF timescale scaled to match the $z=0$ Kennicutt law \citep{ken98a}.
The density threshold for SF is $n_{\rm H}=0.13 \cmc$.
We use a \citet{cha03} initial mass function (IMF) throughout. The
fraction of mass of the IMF going into massive stars, defined here
as $\ge 10 \msolar$ and assumed to result in Type II supernovae
(SNe), is $f_{\rm SN}=0.18$; for reference, a Salpeter IMF
has $f_{\rm SN}=0.10$.

Our simulations directly account for metal enrichment from sources
including Type II supernovae (SNe), Type Ia SNe, and asymptotic
giant branch (AGB) stars.  Gas particles eligible for SF undergo
self-enrichment from Type II SNe using the instantaneous recycling
approximation, where mass, energy, and metallicity are assumed to
immediately return to the ISM.  Type II SN metal enrichment uses
the metallicity-dependent yields calculated from the \citet{chi04}
SNe models.  Our prescriptions for feedback, described below, also
assume the instantaneous input of energy from O and B stars (i.e.
stars that are part of $f_{\rm SN}$).  We input the Type Ia SNe
rates of \citet{sca05}, where an instantaneous component is tied
to the SFR, and a delayed component is proportional to the stellar
mass as described in OD08.  Each Type Ia SN adds $10^{51}$ ergs of
thermal energy and the calculated metal yields of \citet{thi86} to
surrounding gas; the mass returned is small compared to that from
Type II SNe.  AGB stellar enrichment occurs on delayed timescales
from 30 Myrs to 14 Gyrs, using a star particle's age and metallicity
to calculate mass loss rates and metallicity yields as described
in OD08.  AGB mass and metal loss is returned to the three nearest
surrounding gas particles.  OD08 showed that the largest effect of
AGB stars is to replenish the ISM, because a star particle can lose
over 30\% of its mass over $t_{\rm H}$, nearly double the $1-f_{\rm
SN}=18\%$ that is recycled instantaneously via Type II SNe.

Galactic outflows are implemented using a Monte Carlo approach
analogous to star formation.  Outflows are directly tied to the
SFR, using the relation $\dot M_{\rm wind}= \eta \dot M_{\rm SF}$,
where $\eta$ is the {\it mass loading factor}.  The probabilities
for a gas particle to spawn a star particle is calculated as described
above, and the probability to be launched in a wind is $\eta$ times
that.  If the particle is selected to be launched, it is given an
additional velocity of $v_w$ in the direction of {\bf v}$\times${\bf
a}, where {\bf v} and {\bf a} are the particle's instantaneous
velocity and acceleration, respectively; this would create a bipolar
outflow for a thin disk, but in practice the outflow has a much
larger opening angle.  This subgrid model circumvents our inability
to resolve the detailed physics of the ISM that generates winds,
at the cost of introducing additional parameters.  Choices of the
parameters $\eta$ and $v_w$ define the ``wind model".

Once a gas particle is launched, its hydrodynamic (not gravitational)
forces are turned off until either $1.95\times10^{10}/(\vw (\kms))$
years have passed or, more often, the gas particle has reached 10\%
of the SF critical density.  This is intended to simulate the
formation of hot chimneys extending out of a disk galaxy providing
a low resistance avenue for outflows to escape into the IGM, which
the spherically-averaged SPH algorithm at our current resolution
is incapable of resolving.  However, explicit decoupling neglects
energy losses owing to the initial creation of such channels as well as
any drag along their interfaces, and as such cannot drive
turbulence in the ISM.  We note that decoupling has a significant
impact on wind propagation; simulations without such decoupling
explored e.g. by \citet{dal08} show that decoupling greatly enhances
the ability of material to escape a galaxy's ISM.  Hence decoupling
must be regarded as an integral part of the wind models described
in this work, having a significant impact on the results, particularly
those associated with recycling of wind material (which would be
much more prevalent if the hydrodynamic forces were never turned
off).  \citet{spr03b} demonstrated that this decoupling achieves
resolution convergence in the cosmic star formation history.  We
will later explicitly demonstrate resolution convergence for many
of the galaxy properties of interest here.

\begin{table*}
\caption{Simulation runs}
\begin{tabular}{lccccccccccc}
\hline
Name &
$L^a$ &
$\epsilon^b$ &
$m_{\rm SPH}$ &
$m_{\rm dark}$ &
$m_{\rm gal}^c$ &
$\vw$ ($\kms$) & 
$\eta^{d}$ &
$E_{\rm wind}/E_{\rm SN}^{e}$ &
$p_{\rm wind}/p_{\rm *}^{f}$ &
$f_*^g$ &
$f_{\rm wind}^h$
\\
\hline
\multicolumn {4}{c}{} \\
r48n384nw   & 48 & 2.5 & $3.6\times 10^7$ & $1.8\times 10^8$ & $\sim 10^{10}$ & 0 & 0 & 0 & 0 & 0.220 & 0   \\
r48n384vzw  & 48 & 2.5 & $3.6\times 10^7$ & $1.8\times 10^8$ & $1.1\times 10^9$ & $\propto\sigma$ & $\propto\sigma^{-1}$ & 0.49$^i$ & 5.2 & 0.097 & 0.345\\
r48n384cw   & 48 & 2.5 & $3.6\times 10^7$ & $1.8\times 10^8$ & $1.1\times 10^9$ & 680 & 2 & 0.95 & 8.7 & 0.045 & 0.174\\
r48n384sw   & 48 & 2.5 & $3.6\times 10^7$ & $1.8\times 10^8$ & $1.1\times 10^9$ & 340 & 2 & 0.24 & 4.4 & 0.124 & 0.476\\
\hline
r24n256nw$^j$   & 24 & 1.875 & $1.5\times 10^7$ & $7.6\times 10^7$ & $\sim 4\times 10^9$ & 0 & 0 & 0  & 0 & - & 0 \\
r24n256vzw$^j$  & 24 & 1.875 & $1.5\times 10^7$ & $7.6\times 10^7$ & $4.7\times 10^8$ & $\propto\sigma$ & $\propto\sigma^{-1}$ & - & - & - & -\\
r24n256cw$^j$   & 24 & 1.875 & $1.5\times 10^7$ & $7.6\times 10^7$ & $4.7\times 10^8$ & 680 & 2 & 0.95 & 8.7 & - & -\\
r24n256sw$^j$   & 24 & 1.875 & $1.5\times 10^7$ & $7.6\times 10^7$ & $4.7\times 10^8$ & 340 & 2 & 0.24 & 4.4 & - & -\\
\hline
r48n256nw   & 48 & 3.75 & $1.2\times 10^8$ & $6.1\times 10^8$ & $\sim 3\times 10^{10}$ & 0 & 0 & 0  & 0 & 0.182 & 0 \\
r48n256vzw  & 48 & 3.75 & $1.2\times 10^8$ & $6.1\times 10^8$ & $3.7\times 10^9$ & $\propto\sigma$ & $\propto\sigma^{-1}$ & 0.52$^i$ & 5.1 & 0.081 & 0.257 \\
r48n256cw   & 48 & 3.75 & $1.2\times 10^8$ & $6.1\times 10^8$ & $3.7\times 10^9$ & 680 & 2 & 0.95 & 8.7 & 0.046 & 0.179 \\
r48n256sw   & 48 & 3.75 & $1.2\times 10^8$ & $6.1\times 10^8$ & $3.7\times 10^9$ & 340 & 2 & 0.24 & 4.4 & 0.116 & 0.448 \\
\hline
\end{tabular}
\\ 
$^a$ Box size in comoving $\hmpc$.\\
$^b$ Gravitational softening length in comoving $\hkpc$ (equivalent Plummer).\\
$^c$ Galaxy mass resolution limit in $M_\odot$; see discussion in \S\ref{sec:massfcn} of no-wind case.\\
$^d$ $\eta\equiv\dot{M}_{\rm wind}/$SFR.\\
$^e$ Total feedback energy divided by SN energy (Chabrier IMF).\\
$^f$ Total feedback momentum divided by momentum output by $10^7$~yr old stellar population (Chabrier IMF).\\
$^g$ Global mass fraction of baryons in stars at $z=0$.\\
$^h$ Global mass fraction of baryons that have been ejected in winds by $z=0$.\\
$^i$ Volume-averaged to $z=0$; in this model it is $\propto\sigma$.\\
$^j$ Owing to a hardware failure, we only have information at z=0,1,3 for these runs.\\
\label{tab:sims}
\end{table*}

\subsection{Wind Models}\label{sec:windmodels}

For this paper we run four wind models, described as follows:

{\bf No winds (nw):} We turn off our galactic winds.  Note that,
as with all simulations, energy is imparted thermally to ISM SPH
particles using the \citet{spr03a} subgrid two-phase recipe where
all SN energy is instantaneously coupled to the ISM.  Owing to the
explicit coupling of the hot and cold ISM phases, such energy cannot
drive a wind.  

{\bf Constant winds (cw):} Feedback energy is added kinetically to
gas particles at a constant rate relative to the SFR, with $\eta=2$
and a constant velocity $\vw=680 \kms$.  This is intended to mimic
the constant-wind model of \citet{spr03a}, but with some notable
differences.  Here $\eta$ is defined relative to the SFR of the
entire Chabrier IMF, and not just the long-lived stars in a Salpeter
IMF as it is defined by \citet{spr03a}.  Second, $\vw=680 \kms$ is
used rather than $484 \kms$, because there is more SN energy per
mass formed in a Chabrier IMF.  The kinetic energy imparted to wind
particles per unit SF is $9.25\times 10^{48} {\rm erg}/\msolar$,
which is 95\% of the SN energy in this IMF assuming all stars $\ge
10 \msolar$ add $10^{51}$ ergs/SN.  The hydrodynamic decoupling
time in this model has a maximum of $2.9\times 10^7$ years.  We
note that this model is similar to the ``REFERENCE" outflow model
of the OverWhelmingly Large Simulations~\citep[OWLS][]{sch10}, except
that we include hydrodynamic decoupling while they do not; in
fact, they have also considered models likw nw and vzw (below)
except without decoupling.

{\bf Slow winds (sw):} We run an alternative constant-wind model
with the only difference being the outflow velocities are half as
high as the cw winds (i.e. $\vw=340 \kms$ and $\eta=2$).  Therefore
only roughly a quarter of the SN energy couples kinetically into
these outflows.

{\bf Momentum-conserved winds (vzw):} This model uses the analytic
derivations by \citet{mur05} based on the galaxy velocity dispersion
($\sigma$).  Its implementation is explained in \citet{opp06} and
updated in OD08 with the addition of an {\it in situ} group finder
to calculate $\sigma$ using galaxy mass following \citet{mo98}.
The wind parameters vary between galaxies using the following
relations:
\begin{eqnarray}
  \vw &=& 3\sigma \sqrt{f_L-1}, \label{eqn: windspeed} \\
  \eta &=& \frac{\sigma_0}{\sigma} \label{eqn: massload},
\end{eqnarray} 
where $f_L$ is the luminosity factor in units of the galactic
Eddington luminosity (i.e. the critical luminosity necessary to expel
gas from the galaxy potential), and $\sigma_0$ is the normalization of
the mass loading factor.  Here we randomly select $f_L=[1.05,2]$ for
each SPH particle, following observations by \citet{rup05}, and include a
metallicity dependence for $f_L$ owing to more UV photons output by
lower metallicity stellar populations 
\begin{equation}
  f_L = f_{\rm{L},\odot} \times 10^{-0.0029*(\log{Z}+9)^{2.5} + 0.417694}.
  \label{eqn: zfact}
\end{equation}
We further add an extra kick to get out of the potential of the
galaxy in order to simulate continuous pumping of winds until it
is far from the galaxy disk (as argued by MQT05).  These parameters
$\vw$ and $f_L$ are well-constrained by observations, leaving the
only unconstrained free parameter as $\sigma_0$.  We set $\sigma_0=150
\kms$, which is the same efficiency factor of the mass loading used
in OD08.  The energy budget for winds scales as $\propto \eta
v_w^2\propto \sigma$, and exceeds the supernova energy budget only
at quite large halo masses of $\ga 10^{14} M_\odot$ (at $z=0$).
Since we do not include any type of feedback that quenches star
formation in massive galaxies~\citep{gab10}, our simulations do not
accurately model these systems in any case.  In general, the wind
energetics in this model are quite modest, and despite the large
mass outflow rates and speeds around the escape velocity, are
generally well below the supernova energy budget (globally, about half 
in the r48n384vzw run).  This does not even account for radiative
energy input from massive stars, which can exceed the energy output from
supernovae.

These wind models allow us to illustrate how outflows impact the
scaling relations of galaxies.  An important difference comes from
the wind speed scaling:  In the momentum-conserved outflows case,
the wind speed is typically around the galaxy escape velocity, and
scales with it.  Hence in this model, the potential well of the
galaxy does not play a governing role in wind dynamics as a function
of mass.  This stands in contrast to the canonical model for
outflows~\cite[e.g.][]{dek86} in which outflows escape from small
galaxies owing to their small potential wells while being recaptured
in large galaxies.  If gravity was the only relevant force, this
model should show exactly the same escape fraction of outflowing
material from galaxies of all masses.  In actuality, hydrodynamic
slowing plays an important role~\citep{opp08}, and because more
massive galaxies live in denser gaseous environments, this retards
winds more around more massive galaxies, causing greater return of
wind material to these systems~\citep{opp10}.

The constant-$\eta$ models (sw and cw) behave more like the canonical
model for winds, in the sense that at low galaxy masses their
(constant-velocity) winds can escape the potential well, while at
high masses they cannot.  One can envision this as an ``effective"
mass loading factor that is roughly the input value of 2 at small
masses, but drops to zero at large masses~\citep{fin08}.  Alternatively,
one can view this as rapid recycling of wind material back into
galaxies at high masses~\citep{opp10}.  In any case, the constant
wind speed imparts a feature at a mass scale corresponding to where
the assumed wind velocity is comparable to the galaxy escape velocity.
Because of the different wind speeds in cw and sw, this feature
should occur at different masses (roughly a factor of eight apart
in halo mass, given the factor of two difference in wind speeds).
Hence comparing these two models enables us to identify features
in scaling relations that are related to how the effective mass
loading factor scales with mass; this is discussed extensively in
\citet[see e.g. their Figure~4]{opp10}.  As we will argue later,
it is the scaling of the (effective) mass loading factor with $M_*$
that primarily governs most galaxy scaling relations.

Our twelve runs are listed in Table \ref{tab:sims}.  They were run
at the University of Arizona's SGI cluster, ICE, the National Center
for Supercomputing Applications' Intel cluster, Abe, and the
University of Massachusetts' Xeon cluster, Eagle.  Each run took
between 100,000 and 150,000 CPU hours, although nearly 1 million
CPU hours were used when including smaller box sizes and various
test runs.  We note that due to technical difficulties, we only
have galaxy catalogs for the r24 runs at $z=0,1,3$, and do not have
any information regarding satellite galaxies in these runs.  However,
all the important trends can be seen from the other runs at the
redshifts where the r24 runs are not available.

We emphasize that all these models are heuristic and intended only
to explore how these wind scalings impact galaxy properties.  We are
not directly advocating any particular physical process, though we
are attempting to use physically-motivated recipes.  In particular,
the momentum-driven wind scalings can arise from models that only employ
supernova energy to drive winds, owing to the interactions of super-heated
ISM material with surrounding gas~\citep[e.g.][]{dal08}.  Hence the name
``momentum-driven" is only intended to indicate the scalings rather than
a detailed physical model.

Nevertheless, owing to its prominence in this paper and others, we briefly
discuss some issues related to the physics of the momentum-driven scalings
wind model.  The basic physical motivation is a model in which radiation
pressure drives dust outwards, which couples to the gas to generate
an outflow~\citep{mur05,zha10}.  First, it is important to note that
the momentum input required to drive winds at hundreds of km/s with
mass loading factors as assumed above significantly exceeds the photon
momentum emitted by the stellar populations~\citep[M. Haas \& J. Schaye,
priv. comm.][]{haa10}.  We show this ratio in Table~\ref{tab:sims}, and
is a factor of 5.2 for our r48n384vzw simulation\footnote{The momentum
requirements are similarly large for the constant-$v_w$ models, but in
this case the physical motivation is supernova energy, so the momentum
input is not relevant.} Following \citet{haa10}, this is calculated from
the luminosity of stars for a solar-metallicity $10^7$~yr old stellar
population assuming a \citet{cha03} IMF, divided by $c$.  The large
value indicates that momentum alone, under the assumption of single
photon scattering, cannot power the outflows in the vzw model.

It is possible that subsequent re-radiation of photons could cause
significantly more momentum deposition than what is emitted by the
stars if the gas is optically thick to infrared radiation.  For an
infrared optical depth of $\tau_{IR}$, momentum deposition is increased
by $1+\tau_{IR}$.  This factor can only be significantly greater than
unity in highly dense gas where young stars form.  \citet{hop11} computed
$\tau_{IR}$ in simulations of radiation-driven winds from high-resolution
disk galaxies, and indeed found that in some cases $\tau_{IR}$ can
approach an order of magnitude, driving an outflow velocity of $\sim
200$~km/s leaving the disk with mass loading factors of order unity.
While this is encouraging, the issue of whether enough momentum is
available from stars, and whether it can effectively couple to the dust
and in turn to the gas, remains an open question.

Another possibility is that supernovae and photons from young stars work
together to drive winds~\citep[e.g.][]{nat09,mur10}.  Invariably, this
must be happening at some level, since both processes are energetically
comparable.  \citet[Figure 2.19]{haa10} showed that the momentum from
SNe can be an order of magnitude larger than that from stellar radiation
at $10^7$~yr after the burst, although it drops rapidly with time.
\citet{mur10} showed that for a 300~km/s wind, radiation pressure can
yield a mass loading factor up to $\eta=0.4$ in the single-scattering
regime, which is factors of many less than what our momentum-driven wind
scalings require.  Hence supernova momentum input would have to dominate
overall, particularly at large radii where the optical depth to absorb
photons becomes small.  Despite being driven primarily by supernovae and
not radiation pressure, the scalings predicted in this scenario continue
to follow those expected for momentum-driven winds~\citep{mur10}.

In short, the momentum-driven wind scalings that we have assumed
challenge the available momentum budget, but could plausibly arise from
a combination of physical processes.  Much work remains to be done to
marry large-scale constraints from cosmological simulations with detailed
studies of wind expulsion from star-forming regions.

\subsection{Galaxy and halo identification}

We use SKID\footnote{http://www-hpcc.astro.washington.edu/tools/skid.html}
(Spline Kernel Interpolative Denmax) to identify galaxies as bound
groups of star-forming gas and stars~\citep{ker05,opp10}.  Since
we specifically include just star-forming gas, our simulated galaxies
only account for gas within the galactic ISM and not from the
extended halo; this defines the galaxy's gas mass.  Our galaxy
stellar mass limit is set to be $\ge 64$ star particles~\citep{fin06},
resulting in the masses listed under $m_{\rm gal}$ in Table~\ref{tab:sims}.
We will argue that for the no-wind case the resolution convergence
is poorer, and we list approximate masses that show convergence.
Unless otherwise noted, we will only consider galaxies with stellar
mass $M_*\geq m_{\rm gal}$ in our analysis.

We separate galaxies into central and satellite galaxies by associating
each galaxy with a halo.  We identify halos via a spherical overdensity
algorithm~\citep{ker05}.  In brief, we begin at the center of each
galaxy and expand out spherically until the mean density enclosed
equals that of a virialized halo~\citep{kit96}.  We then subsume
all galaxies whose centers are within that of a larger galaxy's
halo into that larger halo.  In this way, each galaxy is a member
of one and only one halo.  The central galaxy is the largest one
that is at the center of the halo, and any other galaxies within
that halo are satellites.

\section{Stellar Mass and Star Formation Rate Functions}\label{sec:msfrfcn}

\subsection{Stellar Mass Function}\label{sec:massfcn}

\begin{figure*}
\vskip -1.0in
\setlength{\epsfxsize}{0.85\textwidth}
\centerline{\epsfbox{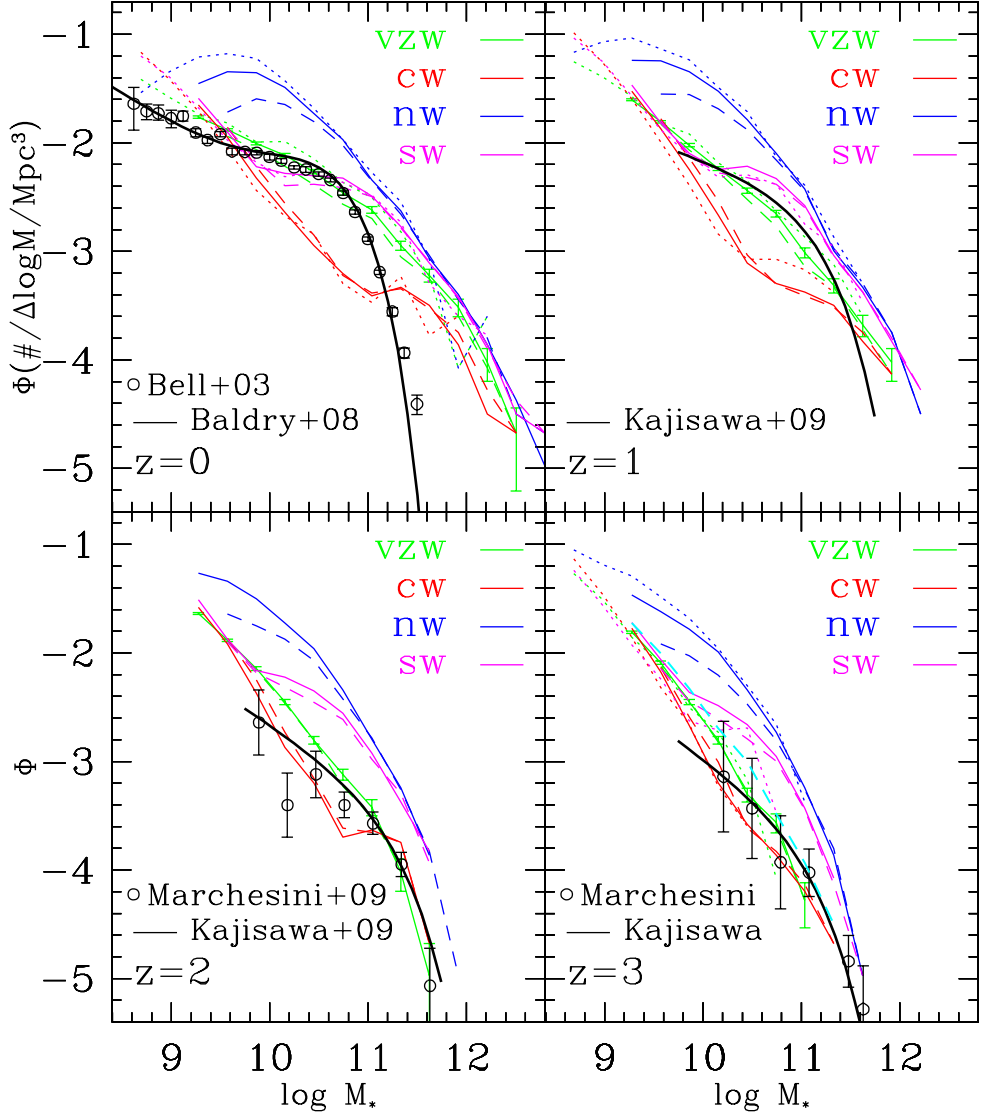}}
\vskip -1.0in
\caption{Stellar mass functions at $z=0,1,2,3$ in cosmological
hydrodynamic simulations employing our four galactic outflow
scalings: momentum-conserved winds (green), constant winds (red),
no winds (blue), and slow winds (magenta).  Solid lines show results
from the r48n384 series of runs, dotted lines show r24n256 series,
and dashed lines show r48n256 series; these illustrate resolution
convergence, although when interpreting convergence one should keep
in mind that these runs are not equally (logarithmically) spaced in
mass resolution: r48n256 is $2.4\times$ r48n384, which is $3.3\times$
r24n256.  The momentum-conserved wind case shows Poisson error bars as
a representative example.  Cyan dashed curve at $z=3$ shows the result
of adding a Gaussian random scatter of $\sigma=0.3$~dex to each galaxy's
stellar mass in the momentum-conserved wind case.  Data points at $z=0$
from \citet{bel03}, curve at $z=0$ from \citet{bal08}, curves at $z=1,2,3$
from \citet{kaj09}, and data points at $z=2,3$ from \citet{mar09}.
}
\label{fig:massfcn}
\end{figure*} 

We begin by considering the stellar mass function and its evolution
in our simulations.  Stellar mass appears to be a governing parameter
for galaxy properties, having fairly tight scaling relations with
many other fundamental galaxy properties, relatively independently
of environement~\citep[e.g.][]{li10}.  Hence the stellar mass
function provides a key barometer for simulated galaxy populations.

Figure~\ref{fig:massfcn} shows the galaxy stellar mass function (GSMF) out
to $z=3$ in our four outflow simulations.  Data is shown for comparison
from \citet[$z\sim 0$, points]{bel03}, \citet[$z\sim 0$, curve]{bal08},
\citet[$z=1,2,3$, curves]{kaj09}, and \citet[$z=2,3$, curves]{mar09}.
These were selected as fairly recent determinations of the stellar mass
function from some of the deepest available near-infrared observations,
though it is far from a complete sample of available data.  The curves
represent Schechter function best fits, while the data points (from
different samples) give an indication of the uncertainties.  Note that
the \citet{kaj09} data plotted at $z=1$ is actually centered at $z\approx
0.75$, but the evolution is not rapid during this period.  Also, we have
scaled all data to a \citet{cha03} IMF (as assumed in the simulations),
by dividing the stellar masses by 1.25 for those quoted using a diet
Salpeter IMF~\citep[namely,][]{bel03,bal08}, and by 1.8 for those quoted
using a Salpeter IMF~\citep{kaj09}.  \citet{mar09} used a \citet{kro01}
IMF which is very similar to Chabrier, so we apply no IMF correction.
Finally, the \citet{bel03} data has been corrected to $h=0.7$; all other
data shown assumed this value for the Hubble constant.

A test of numerical resolution convergence is provided by comparing
the solid, dotted, and dashed lines for each model, which are from
the r48n384, r24n256, and r48n256 series of runs, respectively.
The difference in mass resolution between r48n384 and r48n256 is a factor
of 2.4, while between r48n384 and r24n256 it is 3.3; without accounting
for this, convergence will appear slightly better at low resolution.
In actuality, for the wind models at every redshift the GSMFs show very
good convergence (modulo some stochastic fluctuations at the massive end),
and the faint end of the higher-resolution run joins smoothly onto that
of the larger-volume run.  For the no-wind simulations, this is not the
case: The higher-resolution run deviates from the lower-resolution run
at a $M_*$ that is much larger than the nominal 64 star particle galaxy
mass resolution.  The reason can be traced back to resolution convergence
for halos, as opposed to galaxies:  Halos also require approximately
64 particles to robustly host galaxies, and in the no-wind case this
corresponds to a much higher $M_*$ than in the wind models where star
formation at a given halo mass is much suppressed~\citep{opp10}.
Hence convergence for the no-wind case differs from the wind runs,
and its resolved galaxy mass is higher by roughly an order of magnitude
(as indicated in Table~\ref{tab:sims}).  The low-mass peak of the GSMF
is shifted over by roughly the difference in mass resolution between
the two models, further indicating that the GSMF here is impacted by
resolution effects.  Hence the turn-down in the no-wind GSMF at low
masses is a numerical artifact, not a prediction.  For the wind models,
this turn-down happens below the nominal stellar mass resolution.

The no-wind case (nw; blue), where resolution-converged, shows the
classic behavior that the GSMF well exceeds observations at all
masses~\citep[see e.g.][for a more detailed discussion]{ker09b}.
This is symptomatic of the overcooling problem~\citep{whi91,bal01,dav01},
in which there are too many baryons locked in stars at all epochs
without strong feedback.  We note that this occurs at {\it all}
redshifts and masses back to $z\sim 4$~\citep[and even farther back;
see][]{dav06}, indicating that the overcooling problem is not just
a late-time phenomenon and must be mitigated in the early Universe.

It has long been suggested that galactic outflows could suppress
star formation to solve the overcooling problem.  The magnitude of
suppression, factors of several even at maximum galaxy formation
efficiency around $M^\star$ increasing to much larger factors at
larger and smaller masses, indicates that outflows must eject a
substantial amount of gas from galaxies and/or suppress accretion
by adding energy to the surrounding IGM.  In star-forming galaxies
(typically $M_*\la M^\star$), the dominant energy source is star
formation, either from supernovae or young stars.  This motivates
all of our wind models having mass loading factors exceeding unity.
For concreteness, Table~\ref{tab:sims} shows as its final two columns
the global mass fraction of baryons in stars at $z=0$ and ejected
in winds by $z=0$, respectively.  For the momentum-conserved run,
the ratio between the two is roughly 3.5, i.e. $3.5\times$ more
baryons have been ejected from galaxies than have formed into stars
by $z=0$.  The constant-$\eta$ models have a ratio of approximately
3.8.  That this ratio signficantly exceeds unity is a generic feature
of all models that solve overcooling via an ejective feedback
mechanism such as galactic outflows.

The exact shape of the $z=0$ GSMF depends sensitively on outflows,
albeit in a subtle way.  As discussed in \citet{opp10}, it is not
the overall suppression by ejection that governs the shape, but
rather the effectiveness of subsequent re-accretion of ejected
material, i.e. recycled wind mode accretion.  As shown in \citet{opp10},
without wind recycling, all GSMFs look fairly similar, and show a
steep power law behavior with a faint-end slope comparable to that
of the halo mass function.  Because wind material re-accretes into
larger galaxies faster (``differential recycling"), it boosts
high-mass galaxy masses relative to low-mass ones, and therefore
flattens the faint end of the stellar mass function.  At sufficiently
small masses, the timescale for wind re-accretion exceeds the Hubble
time, at which point the mass function returns towards tracking the
dark matter halo mass function as there is no differential effect
from recycling.  At sufficiently large masses, recycling is so fast
that it effectively is like having no winds at all.  This creates
a three-tier behavior for the GSMF in simulations: It follows the
steep dark matter mass function both above and below the mass range
where differential recycling operates, and it is flatter within
that mass range.

All our wind models display this three-tiered behavior, though it
is less evident in some cases.  At high masses, they have a slope
similar to that of the no-wind model.  In the intermediate mass
regime, the slope is flatter or even inverted, and finally at low
masses once again the GSMF is reverts to being steep.  We note that
all models grossly overpredict the massive end of the GSMF compared
to observations.  This is because we do not have any feedback
mechanism to truncate star formation in massive galaxies, such as
feedback from active galactic nuclei~\citep[AGN; e.g.][]{dim05}.
Hence our first key point is that our current simulations can only
reliably probe the regime of star-forming galaxies, which dominate
at masses $\la M^\star\approx 10^{11}M_\odot$~\citep{sal07}.  In
\citet{gab10} we show that it is possible to include empirical
feedback mechanisms to quench star formation in massive galaxies,
without significantly affecting the population of galaxies below
$M^*$.  This is consistent with the observation that strong AGN
activity is not seen in lower-mass galaxies today~\citep[$M_*\la
10^{10.5}M_\odot$;][]{kau04}.

The exact transitions between the various GSMF tiers depend strongly
on wind model.  In the constant-$\eta$ cases, the three-tiered
behavior is manifested as a bump in the GSMF at high masses at
$z=0$.  In the constant wind case (cw; red), the middle tier covers
a small mass range, and differential recycling is so strong that
it actually produces an inverted slope of the GSMF within that small
mass range.  Outside that, the GSMF is quite steep, roughly following
the slope of the no-wind case.  The slow-wind case (sw; magenta)
follows a similar pattern, but the mass scale of the differential
recycling is lower because wind speeds are slower, and hence wind
recycling becomes effective at a smaller scale.  In neither case
does the GSMF resemble the observed one at $z=0$ even in the regime
where quenching feedback is unimportant, although the slow-wind
case is not as far off.

The momentum-driven scalings case (vzw; green) produces a more gradual
differential recycling curve than the constant-$\eta$ cases~\citep{opp10}.
Hence the middle tier is not inverted, but merely shallower than
the high and low mass ends of the GSMF.  This mimics the behavior
exhibited by the observed GSMF.  Recent observations by \citet{bal08}
have conclusively detected an upturn to a steep faint-end slope.
In detail, the vzw simulation is still mildly steeper than observed
and the upturn occurs at slightly too high a mass.  If our physical
interpretation for the origin of the GSMF is correct, this would
imply that the momentum-conserved wind model has a differential
recycling curve that is close to, but not quite, correct.  It is
worth mentioning that modeling the dynamics of wind recycling is
subject to significant numerical difficulties as discussed in
\citet[][though more so at in higher-mass halos where hot gas is
present]{opp10}, so the level of discrepancy with data is probably
within modeling uncertainties, but more careful simulations are
needed to determine this.

The evolution of the GSMF can now be probed to high redshifts thanks
to advancing deep near-IR surveys.  At the massive end, the observed
GSMF has a less pronounced cutoff at high-$z$.  Since this truncation
is associated with the presence of ``red and dead" galaxies, the
relative dearth of such galaxies at $z\ga 2$ explains why the
truncation is much less sharp.  At $z\sim 2-3$, the vzw run yields
a GSMF at $M_*\ga 10^{10.5}M_\odot$ that is in fair agreement with
data, although it appears somewhat too steep.  Meanwhile, sw
overproduces the number of galaxies at these stellar masses similar
to the no-wind case.  The constant wind case generally matches well
here, although it still shows a bump at $M_*\ga 10^{11} M_\odot$
that is not seen in the data.  Considering all redshift from $z=0-3$
in toto, the momentum-driven scalings model yields the best match
from amongst these models to the population of star-forming galaxies
at $10^{10}\la M_*\la 10^{11}M_\odot$.

The faint-end slope of the GSMF has been observed to evolve towards
being steeper at high redshifts.   Our simulations follow this trend
qualitatively, but only within the intermediate mass regime where
differential recycling flattens (or inverts) the GSMF slope at later
times.  Outside of this regime, the slope always approaches the
faint-end halo mass slope of $\alpha\approx 2$.  For instance, in
the vzw run, the intermediate mass regime occurs at $10^{9.7}\la
M_*\la 10^{10.7} M_\odot$, and the slope at $z=0$ is $\alpha\approx
1.4$, which is close to the observed slope in that mass range.  In
the sw case, the intermediate mass regime is narrower, $10^{10}\la
M_*\la 10^{10.5} M_\odot$, while in the cw case it is equally narrow
but occurs at a mass $\sim 8\times$ higher.  In both constant-$\eta$
simulations, the GSMF slope at $z=0$ is inverted in this regime,
i.e. $\alpha<1$, but at higher redshifts $\alpha$ increases.

The general reason for the increase in $\alpha$ within the intermediate
mass regime is that wind recycling becomes less important at higher
redshifts.  This is primarily because there is less time for ejected
material to return to galaxies~\citep{opp10}.  Our simulations,
particularly the momentum-conserved case, tend to find median
recycling times that are fairly constant at $\sim 1-2$~Gyr, independent
of cosmic epoch~\citep{opp08}; this is partly because wind speeds
at a given $M_*$ are higher~\citep{opp08} at high-$z$.  Hence the
features associated with recycling become less prominent, and in
fact they are essentially invisible at $z\ga 2$ in the vzw case.

Observationally, $\alpha$ is seen to evolve as $\alpha\approx
\alpha_0+(0.082\pm0.033)z$~\citep{fon06}, where $\alpha_0\approx
1.2-1.3$.  Effectively, this corresponds to the slope in the mass
regime of $10^{10}\la M_*\la 10^{11}M_\odot$, which is what can be
probed in practice out to high-$z$ at this time; at lower masses,
the present-day GSMF becomes steeper, but this cannot be reliably
traced at high-$z$.  The momentum-driven scalings model matches
well at low-$z$ (below $M^\star$), but at high-$z$ the faint end
is quite a bit steeper: e.g. at $z=2$, $\alpha\approx 2$ in the
regime where it is observed to be more like 1.5 or so.  Hence while
the evolutionary trend is qualitatively as observed, there are
quantitative discrepancies in the sense that this model has too
many small galaxies (as is the case with all models).  This has
also been noted at $z\sim 4-7$ in a comparison of various simulations
to mass function observations~\citep{gon11}; all models have too
steep a mass function at $M_*\la 10^9 M_\odot$.  This may indicate
that feedback processes in very small systems are not being correctly
represented in these models, a point that will be reiterated
throughout this paper.

One issue that may explain part of the difference between the simulations
and data are observational uncertainties in determinations of stellar
masses, which can become significant particularly at high-$z$.
To explore this, at $z=3$ we added a Gaussian random scatter with
$\sigma=0.3$~dex to each galaxy's stellar mass in the momentum-conserved
wind simulation~\citep[which is a typical observational uncertainty;
see][]{mar09}, and recomputed the stellar mass function.  The result is
shown as the dashed cyan line in Figure~\ref{fig:massfcn}.  Because the
mass function is steep, more small galaxies get scattered to larger masses
than vice versa, and the mass function does becomes slightly shallower
(and in this case, agrees slightly better with observations at the massive
end).  But this does not go far towards reducing the overproduction of
small galaxies in the simulations.  There are, additionally, substantial
systematic uncertainties in determining stellar masses from photometric
data~\citep[e.g.][]{mar09}; we have not considered such effects here,
but it's conceivable that they could dominate over the statistical errors.
Hence while this effect should be accounted for when comparing carefully
to data, it does not qualitatively alter the faint end discrepancy.

Overall, the momentum-driven scalings model does the best job of
the models considered in reproducing the observed GSMFs around
$M^\star$ from $z=3\rightarrow 0$.  There are still significant
discrepancies at higher and lower masses, so the agreement is 
only good in the range of $10^{10}\la M_*\la 10^{11}M_\odot$.
Nevertheless, this is a significant success, a first for cosmological
hydrodynamic simulations (that we are aware of).  While the discrepancy
at high masses is well-known and strongly suggestive of quenching
feedback absent in these simulations, the discrepancy at small
masses is less certain owing to still substantial observational
uncertainties at high-$z$.  Upcoming deeper observations should
quantify this discrepancy to greater precision, possibly providing
insights into the nature of feedback processes in high-redshift
dwarf galaxies.

\subsection{Star Formation Rate Function}\label{sec:sfrfcn}

\begin{figure*}
\vskip -1.0in
\setlength{\epsfxsize}{0.85\textwidth}
\centerline{\epsfbox{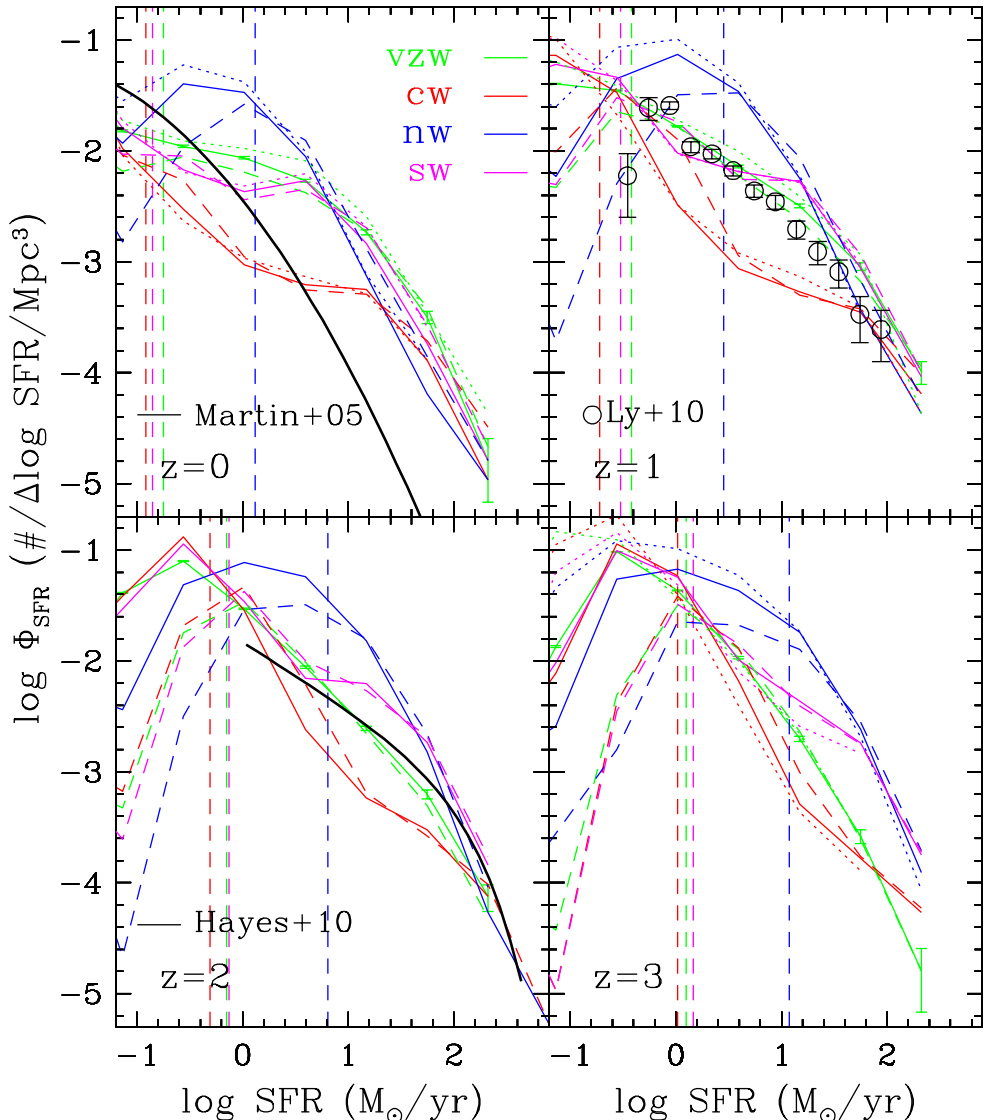}}
\vskip -1.0in
\caption{Star formation rate functions $\Phi_{\rm SFR}$ at $z=0,1,2,3$
in our r48n384 series of simulations employing our four galactic outflow
scalings: momentum-conserved winds (green), constant winds (red), no
winds (blue), and slow winds (magenta).  Solid lines show results from
the r48n384 series of runs, dotted lines show r24n256 series ($3.3\times$
better mass resolution), and dashed lines show r48n256 series ($2.4\times$
worse mass resolution).  The momentum-conserved case shows Poisson error
bars as a representative example.  Dotted lines indicate the SFR at
galaxy stellar mass limit (see Table~\ref{tab:sims}) at each redshift,
from a linear fit to the $M_*-$SFR relation for each model; below this
SFR, the results are less likely to be robust.  The thick line at $z=0$
shows $\Phi_{\rm SFR}$ derived from ultraviolet plus infrared data
from \citet{dmar05}, while points at $z=1$ and thick line at $z=2$ show
$\Phi_{\rm SFR}$ derived from observed H$\alpha$ luminosity functions at
$z\approx 0.81$ by \citet{ly10} and at $z\approx 2.2$ by \citet{hay10},
respectively.
}
\label{fig:sfrfcn}
\end{figure*} 

Figure~\ref{fig:sfrfcn} shows the star formation rate function
($\Phi_{\rm SFR}$) for galaxies in our four wind models at four
redshifts.  The dotted line shows the SFR at the galaxy stellar
mass limit listed in Table~\ref{tab:sims}, obtained by fitting the
relationship between SFR and $M_*$ as displayed in Figure~\ref{fig:ssfr}.
Since we do not have a clean stellar mass threshold in this plot,
we show all galaxies, but $\Phi_{\rm SFR}$ at SFRs below the dotted
line may not be robust.  Because the relationship between SFR and
$M_*$ varies slightly with wind model, each model has a different
SFR threshold.  Note that the no-wind case has a significantly
higher resolved galaxy mass threshold in Table~\ref{tab:sims} for
reasons described in the previous section, and hence its dotted
line is at a much higher mass.  Observations at $z=0$ from a
GALEX+IRAS compilation by \citet{dmar05} are shown as the solid
black line, at $z\sim 1$ from \citet{ly10}, and at $z=2$ from an
H$\alpha$ survey by \citet{hay10}.

At $z=0$, $\Phi_{\rm SFR}$ shows different behaviors amongst the various
wind models that mimic many features seen in the GSMF.  The effect
of differential recycling is seen in the three-tiered behavior of the
constant-$\eta$ models, but the intermediate regime is not obviously
seen in the vzw case.  Nevertheless, relative to no-winds, vzw shifts
star formation from smaller to larger systems, owing both to its mass
loading dependence as well as differential recycling.  The large amount of
recycled wind accretion at late times~\citep{opp10} produces a $\Phi_{\rm
SFR}$ at high masses that is actually higher for all the wind models
than for the no-wind case.  The winds are metal-enriched and so are
particularly effective at cooling back onto the galaxy (often failing
to shock heat to the temperature of surrounding halo gas despite their
large outflow velocities), and they delay star formation towards later
epochs making the effect at high masses stronger towards lower redshifts.

Moving to higher redshifts, the strength of the differential recycling
features in the cw and sw simulations lessen as with the GSMF, but
they are still prominent even at the highest redshifts.  For vzw,
$\Phi_{\rm SFR}$ becomes quite a bit steeper at high-$z$.  The shape
of the no-wind $\Phi_{\rm SFR}$ does not vary, but the amplitude
evolution of $\Phi_{\rm SFR}$ is fairly constant from $z=3\rightarrow
2$, and then drops with time reflecting the overall drop in cosmic
SFR.  For example, $\Phi_{\rm SFR}(10 M_\odot/$yr) at $z\sim 2-3$
is about $\sim 10^{-1.5}$, dropping to $\sim 10^{-2}$ at $z=1$ and
$10^{-3}$ at $z=0$.  In contrast, the momentum-conserved wind model
shows a less rapidly evolving $\Phi_{\rm SFR}$, as $\Phi_{\rm SFR}(10
M_\odot/$yr)$\approx -2.3$ at all redshifts, although $\Phi_{\rm
SFR}$ shows a flatter faint-end slope at low redshifts.

Compared to observations at $z=0$~\citep{dmar05}, none of the models
fare well, as they all overproduce the number of star-forming
galaxies at higher SFR's.  This is not surprising, and can be
straightforwardly traced to the lack of any feedback mechanism to
quench star formation in massive galaxies~\citep{gab10} as observed
in the real Universe~\citep[e.g.][]{kau04}.  Note that the number
of star-forming galaxies must be suppressed down to quite moderately
star-forming systems.  For instance, in the momentum-conserved case,
$\Phi_{\rm SFR}$ is overproduced at SFR$\ga 1 M_\odot$/yr.  Therefore
galaxies larger than the Milky Way must be increasingly quenched,
which coincides with the discrepancy in the GSMF above $M^\star$.
The star formation rate function therefore provides a strong
constraint on models for quenching massive galaxies.

At higher redshifts, we compare to the H$\alpha$ luminosity function
determined at $z=0.81$ by \citet{ly10} and at $z\approx 2.2$ by
\citet{hay10}, both of which probe down to $\sim 1 M_\odot/$yr.  We
obtain SFR from $L_{H\alpha}$ using the \citet{ken98b} conversion
divided by $1.8$ to correct from Salpeter to Chabrier IMF, namely
SFR$=4.4\times 10^{-42} L_{H\alpha} M_\odot/$yr.  For \citet{ly10}
we show their extinction- and incompleteness-corrected 2.5$\sigma$
sample.  For \citet{hay10} we show their ``combined" best-fit
Schechter function which is supplemented by bright-end data from
\citet{gea08}, and correct for 0.977 magnitudes of extinction
($A_V=1.19$ at 6563\AA) independent of $L_{H\alpha}$.  At these
epochs, quenched massive galaxies are increasingly rare, so the
comparison should be more meaningful than at $z=0$.

At both epochs, the no-winds run overproduces the number of galaxies
with SFR$\la 100 M_\odot$/yr, sw overproduces galaxies with SFR$\ga
1 M_\odot/$yr, and cw underproduces galaxies with $1\la$SFR$\la 100
M_\odot/$yr.  The momentum-driven scalings model fares reasonably
well at both redshifts.  At $z\sim 1$ it still slightly overproduces
high-SFR galaxies, which is the onset of a trend that becomes much
more prominent by $z=0$.  At $z\sim 2$ the bright end matches well,
while the faint end is modestly steeper than observed.  Note however
that \citet{hay10} derive a steeper H$\alpha$ luminosity function
from their faint-end data alone ($\alpha=1.72$ instead of $\alpha=1.60$
for the fit shown here).  Furthermore, convolving in the uncertainties
in $M_*$ determinations will tend to make the predicted $\Phi_{\rm
SFR}$ shallower, as we showed with the GSMF in Figure~\ref{fig:massfcn}.
Hence the discrepancies may not be significant.

Overall, the simulated star formation rate functions of the
constant-$\eta$ models display a three-tiered trend arising from
differential wind recycling as seen in the GSMF.  The form and
evolution of $\Phi_{\rm SFR}$ provides independent constraints on
how outflows govern star formation across cosmic time.  At low-$z$,
all current models produce too many rapidly star-forming galaxies,
owing to a lack of quenching feedback in massive galaxies.  At $z\ga
1$, the momentum-driven scalings simulation predicts $\Phi_{\rm
SFR}$ in good agreement with available H$\alpha$ data, while other
wind models do not.

\section{Specific Star Formation Rate}\label{sec:ssfr}

The specific star formation rate (sSFR) measures the intensity of
ongoing versus past-averaged star formation, so it provides an
important barometer for how a given galaxy has assembled.  It also
most directly governs the observed colors of galaxies.  The main
sequence of galaxies reflects a relation between sSFR and $M_*$,
and yields important insights into galaxy growth.  In this section
we study the specific star formation rate in our simulations.  We
will particularly focus on how the slowly-evolving equilibrium
between inflows, star formation, and outflows governs the behavior
of sSFR as a function of mass and epoch.

\subsection{Specific Star Formation Rate vs. $M_*$}\label{sec:sfrmstar}

\begin{figure*}
\vskip -1.0in
\setlength{\epsfxsize}{0.85\textwidth}
\centerline{\epsfbox{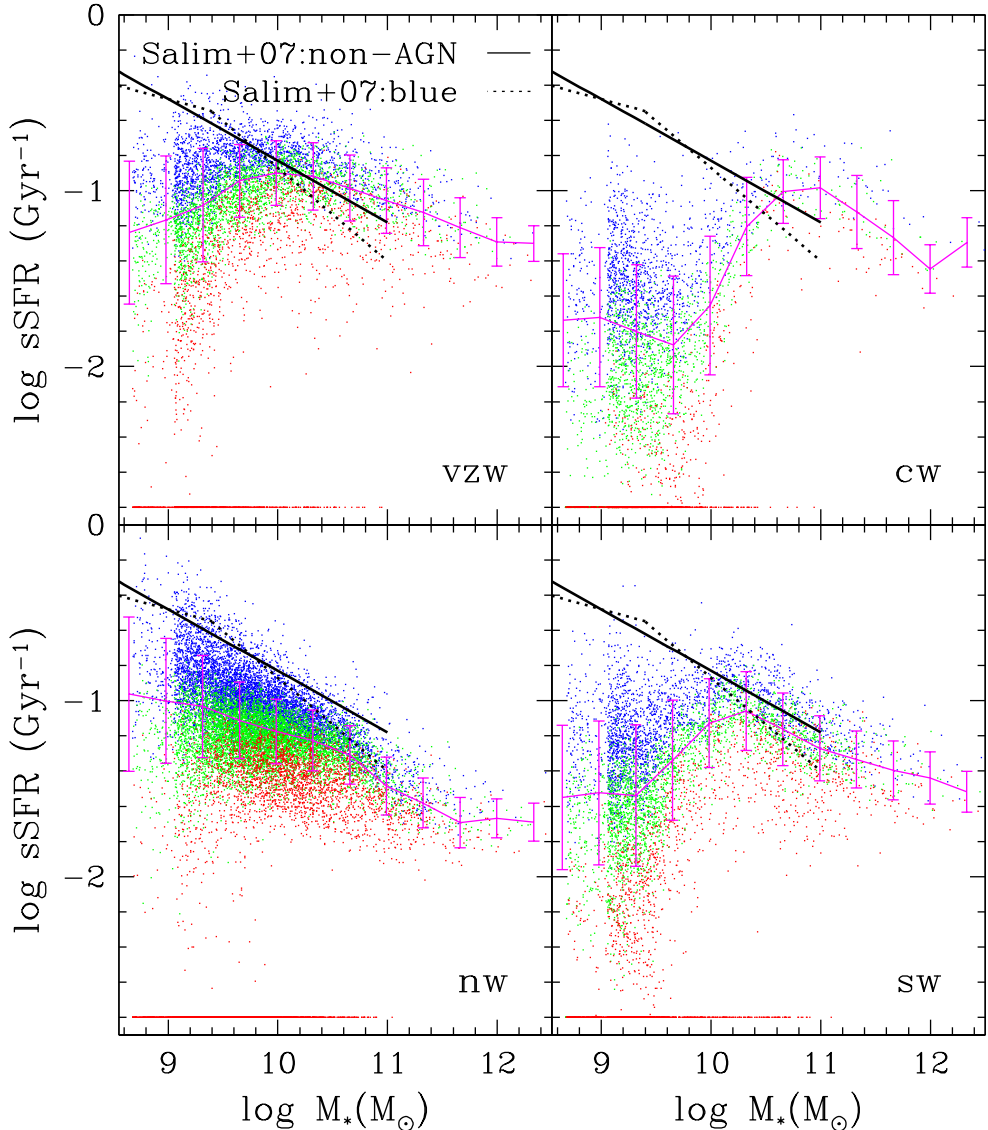}}
\vskip -1.0in
\caption{Specific star formation rate (sSFR$\equiv$SFR$/M_*$) as a function of
stellar mass
at $z=0$ in our r48n384 series of simulations employing our
four galactic outflow scalings: momentum-conserved winds (upper left),
constant winds (upper right), no winds (lower left), and slow winds
(lower right).  Points are color-coded by gas fraction within
each stellar mass bin: Galaxies with $\fgas$ exceeding $0.5\sigma$ above
the median are blue, less than $0.5\sigma$ below the median are red,
and in between are green.  Galaxies with sSFR$=0$ are shown as red
points along the bottom.  A running median is shown in magenta,
excluding non-starforming galaxies.  Thick lines show 
observed GALEX+SDSS best-fit relations~\citep{sal07} for non-AGN 
star-forming galaxies 
(solid) and blue galaxies with $NUV-r<4$ (dashed, broken power law).
}
\label{fig:ssfr}
\end{figure*} 

Figure~\ref{fig:ssfr} shows the sSFR of galaxies at $z=0$ in our
four wind models.  A running median within mass bins is shown in
magenta with $1\sigma$ variance; this does not include the
non-starforming galaxies shown along the bottom of the plot, which
do not substantially change the median except at the lowest masses
in the constant-$\eta$ models (we will return to this in \S\ref{sec:sat}).
The points are color-coded by gas fraction within each mass bin,
blue meaning high $\fgas$ ($>0.5\sigma$ above median), red for low,
and green in between.  Here we define gas fraction as $f_{\rm
gas}\equiv M_{\rm gas}/(M_{\rm gas}+M_*)$, with $M_{\rm gas}$ including
all gaseous phases in the star-forming ISM.  Observations of galaxies
are shown as the thick lines from a compilation of GALEX and SDSS
data by \citet{sal07}: Solid line represents the median sSFR for
all galaxies classified as having no AGN content in a \citet{bal81}
diagram, while the dashed double power-law represents those classified
as red (NUV$-r>4$).  The observed median for all galaxies drops
sharply above $\ga 10^{11}M_\odot$ as red and dead galaxies dominate
in that mass range.

Simulations generically produce a fairly tight relation between SFR
and $M_*$~\citep[e.g.][]{dav00,fin06}.  In all cases, the relatively
flat sSFR($M_*$) implies that star formation rate scales roughly
linearly with stellar mass, and at the high-mass end in all cases
the scaling is sub-linear.  The basic similarity indicates that the
sSFR is not highly sensitive to feedback, as has been noted
previously~\citep{dav08,dut10}.  The basic reason is that, to first
order, feedback processes reduce SFR and $M_*$ in conjunction,
thereby keeping sSFR roughly similar.  All models produce massive
galaxies that are forming stars vigorously, in contrast with
observations that indicate most galaxies at high masses are passive,
indicating the need for some quenching mechanism~\citep{gab11}.
All models also produce a population of low-mass galaxies that are
not forming stars; these are predominantly satellite galaxies, and
will be discussed in \S\ref{sec:sat}.

The basic trends in this plot can be understood in the context of
a balance between accretion, outflow, and star formation.  As shown
in \citet{fin08}, galaxies live in a slowly-evolving quasi-equilibrium
between these quantities, such that 
\begin{equation}\label{eqn:equil}
\dot{M}_{\rm inflow} = \dot{M}_{\rm outflow} + {\rm SFR}, 
\end{equation}
where the first two terms are the mass inflow and outflow rates,
respectively.  We can rewrite this using the definition of the mass
loading factor $\eta\equiv \dot{M}_{\rm outflow}/$SFR, so that
\begin{equation}\label{eqn:sfr} 
{\rm SFR} = \dot{M}_{\rm inflow}/(1+\eta).
\end{equation} 
This equation, while simple, implicitly makes some non-trivial
assumptions.  First, it assumes that the variations in the inflow rate
are slow compared to the rate of processing gas into stars, so that
galaxies are never strongly out of equilibrium (e.g. this is not valid
during major merger events).  Second, it assumes that galaxies do not
collect large reservoirs of gas and then process it, but rather process
gas as it is made available to the ISM (from inflows and outflows); we
call this ``supply-regulated" star formation, to distinguish it from
other scenarios where gas aggregates in or around galaxies before being
consumed rapidly~\citep[e.g.][]{egg62,mar10}.  Third, there is no explicit
dependence on cosmic epoch or environment, as these are assumed to be of
secondary importance in galaxy growth.  Fourth, it neglects additional
reservoirs of gas such as stellar mass loss~\citep{sch10,lei10} and wind
recycling~\citep{opp10} that are particularly important at late times,
and hence this equation is expected to be more valid at earlier epochs.
We will show in this paper and Paper~II~\citep[see also][]{fin08} that
simulated galaxy properties generally follow those expected from the
equilibrium relation (eq.~\ref{eqn:equil}), although moreso at early
epochs, suggesting that at least in these models, the above assumptions
are broadly valid.

Within the context of this equilibrium model, let us return to
examining sSFRs.  In the no-wind case, $\eta=0$, so the trend of
sSFR($M_*$) reflects the inflow rate into galaxies of a given $M_*$.
Owing to the growth of hot gaseous halos that retard inflow in
larger halos~\citep[e.g.][]{ker05,dek06}, the sSFR is smaller at
larger masses.  This is one manifestation of downsizing, namely
star formation rate downsizing as defined in \citet{fon09}.  This
model also produces archaeological downsizing, which means that
larger galaxies have older stellage populations~\citep[also called
``natural downsizing"][]{dav06b,nei06}, as we will show in
\S\ref{sec:age}.  Note that these downsizing trends, while qualitatively
in agreement with data, do not by themselves produce massive red
and dead galaxies~\citep{gab10}.  In the no-wind case, the slope
of sSFR($M_*$) agrees well with observations, but the amplitude is
too low by $\sim\times 2$.  Hence paradoxically, while no winds
produce far too much global star formation (and stellar mass), they
produce too little star formation at a given stellar mass today.
They also produce $\sim\times 10$ more stellar mass for a given
halo mass~\citep{ker09a,opp10}, so that the excess global star
formation arises from too much star formation per halo, particularly
at high and low masses.

The wind models deviate from the no wind case based on $\eta(M_*)$
and wind recycling.  In the cw and sw models, since $\eta$ is constant,
in the simplest scenario both SFR and $M_*$ are reduced equally.  This
shifts a given galaxy down in $M_*$ from the no-wind case by a factor
of $(1+\eta)=3$, when the mass is small enough that winds don't recycle
(i.e. the effective mass loading factor is identical to the input mass
loading factor).  Since the overall slope of the no-wind case is negative,
this results in a lower sSFR at a given mass.  An additional suppression
is provided by preventive feedback from heat added to surrounding gas by
winds~\citep{vdvoort10}, which lowers the accretion rate and hence sSFR at
a given $M_*$.  This is particularly noticeable in the cw case which adds
substantial energy to surrounding gas and the even the IGM~\citep{opp06}.
At the highest masses, all winds are gravitationally and hydrodynamically
trapped, resulting in an effective mass loading factor of $\eta\approx 0$
as discussed in \S\ref{sec:windmodels}, and so also follows the no-wind
case trend with a slightly higher amplitude.  The reason for the amplitude
difference is that wind recycling removes mass from small systems and
adds it back into larger systems at later epochs.  Hence wind recycling
shifts star formation from low to high masses, an effect that is stronger
with the higher wind speeds owing to the increase in the recycling time
and increasing preventive feedback from added heat.  Both these models
are in poor agreement with observations, as they have too low sSFR by
up to an order of magnitude at the low-mass end, and the observations
do not indicate any positive-slope feature in the predicted mass range.

In the case of momentum-conserved winds, $\eta$ varies with mass.
At the lowest masses where there is no wind recycling, $\eta\propto
\sigma^{-1}\propto M_{\rm gal}^{-1/3}\propto M_*^{-1/3}$ (roughly,
although in detail smaller galaxies tend to have lower stellar fractions,
making the slope slightly shallower), since our wind model ties $\sigma$
directly to the galaxy baryonic mass $M_{\rm gal}$ as identified by
our group finder~\citep[see eq.~6 of][]{opp08}.  Hence the sSFR shows
a slope that exceeds the no-wind case by roughly $1/3$.  This by itself
would results in a slope of around zero.  But differential wind recycling
produces an additional effect, which in this model is more gradual and
occurs over a wider range in mass than in the constant-$\eta$ cases.  At
intermediate masses, the additional fuel from recycling causes a positive
slope for sSFR($M_*$), for the same reason as in the constant-$\eta$
case except the effect is more gradual.  \citet{fir10} also pointed
out that wind recycling tends to increase the slope of sSFR($M_*$).
At the largest masses, the slope again reverts to the no-wind case as
recycling times are very short~\citep{opp10}, so winds have little effect.

The gradual rolling of the slope of sSFR($M_*$) is also seen in
observations, but it less pronounced and really only evident at
masses below what are plotted here~\citep[see Figure~16 of][]{sal07}.
Even at $M_*\la 10^9 M_\odot$, the observed sSFR($M_*$) has a
negative slope~\citep{sal07}, whereas this model predicts a positive
slope.  We will see in Paper~II that this is also reflected in the
relation between gas fraction and $M_*$, which has a turnover to
low masses that is likewise absent in observations.  This could
signify that wind recycling should be more effective in small
systems, or perhaps less material should be ejected from these
systems in the first place.  Just as with the faint end of the
high-redshift GSMF, this highlights that even our most successful
model has difficulty correctly reproducing observations of dwarf
galaxies.  We will return to this point in \S\ref{sec:age}.

An interesting second-parameter trend is indicated by the color of
the simulated data points.  Galaxies with high gas fractions at a
given $M_*$ lie above the mean sSFR relation, and those with low
gas fractions lie below.  This is not surprising since the presence
of gas drives star formation based on our assumed star formation
law, but it provides an interesting generic prediction (independent
of outflows) of this class of models where star formation is driven
by accretion.  Such a trend would not be expected, for example, if
mergers drove elevated SFRs at a given $M*$; in that case, it is
the gas configuration rather than overall gas content that drives
elevated SFRs, as gas is much more concentrated in mergers owing
to tidal dissipation~\citep[e.g.][]{mih96}.  In Paper~II we will
discuss second-parameter dependences of the relationships between
stellar mass, metallicity, and star formation rate in much greater
depth, all within the context of the aforementioned equilibrium
model (eq.~\ref{eqn:equil}) for galaxy growth.

\subsection{Specific Star Formation Rate Evolution}\label{sec:ssfrevol}

\begin{figure}
\setlength{\epsfxsize}{0.65\textwidth}
\centerline{\epsfbox{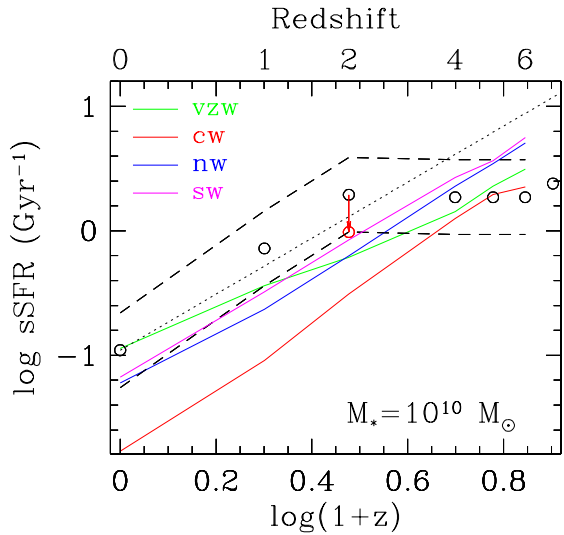}}
\vskip -3.2in
\caption{Evolution of the specific star formation rate
at a stellar mass of $M_*=10^{10}M_\odot$
from $z=6\rightarrow 0$ in our r48n384 series of simulations
employing our four galactic outflow scalings.  Observations are
shown at $z=0,1$ from \citet{elb07} and at $z\geq 2$ from \citet{gon10}.
The downward arrow on the data point at $z\sim 2$ indicates the
$\sim\times 2$ lowering of the UV-derived SFR based on {\it Herschel}
data.  Dashed lines show the typical observed $1\sigma$ variance
about the $M_*-$SFR relation~\citep{noe07,dad07}.  The black dotted
line shows a power-law scaling of $(1+z)^{2.25}$ as predicted
from cold accretion-driven galaxy growth.
}
\label{fig:ssfrevol}
\end{figure} 

Figure~\ref{fig:ssfrevol} shows the evolution from $z=6\rightarrow
0$ of the specific star formation rate at a particular stellar mass
of $M_*=10^{10}M_\odot$ in our four wind models.  This mass is
chosen because observations can directly measure sSFR at this mass
at all epochs (with small extrapolations), and it is well-resolved
by simulations but is expected to be negligibly affected by quenching
feedback.  In the simulations, this sSFR is computed as the median
sSFR within a stellar mass bin of $\pm$0.25~dex around $10^{10}M_\odot$.

Most models show a characteristic evolution of the sSFR which is
essentially a power law in $(1+z)$.  The fundamental physics giving
rise to this is gas inflow driven by growth of structure.  In the
cold accretion paradigm that captures the basic behavior in
hydrodynamic simulations~\citep[e.g.][]{ker05}, the amount of gas
inflowing into the star forming region is correlated with the gas
inflowing at the halo virial radius, since cold streams efficiently
channel material to the center of the halo~\citep{dek09}.  Feedback
processes that strongly impact the surrounding gas to retard infall
can cause significant deviations from this~\citep{vdvoort10}, but
let us set that aside for now to examine the implications of this
simplest scenario.  The amount of gas entering the virial radius
can be estimated by the total mass accretion rate times the baryon
fraction.  The total mass accretion rate onto halos has been well
measured in simulations~\citep{dek09}, and scales with redshift as
$(1+z)^{2.25}$.

The scaling of $(1+z)^{2.25}$ is indicated by the dotted line,
arbitrarily normalised to the observed $z=0$ sSFR.  The no-wind and
slow-wind cases follow this power law essentially at all redshifts.
The no-wind case is easy to understand, since there is no additional
physics regulating star formation beyond accretion.  The slow-wind
case also follows this because it simply ejects a constant two-thirds
of the accreted material.  Hence the SFR reduction translates into
a similar reduction in $M_*$, so the sSFR evolution is nearly
identical to the no-wind case.  In the sw case, two additional
processes are at work:  Feedback heats surrounding gas to retard
accretion into the ISM, which is countered by additional accretion
provided by recycled winds.  Note that at this stellar mass, about
half the $z=0$ accretion is recycled winds~\citep{opp10}, and both
recycling and heating increase to low redshifts.  The impact of
these processes does not appear to be large (barring a remarkable
cancellation), as the net effect is only a slight increase in sSFR
in the sw model over the no-wind case.

The constant wind case shows a substantial departure from no-winds,
with sSFR lower by $\sim\times 2-5$, increasing at late times.  As
discussed previously, this model produces significant heating of
surrounding gas, substantially retarding inflow into the ISM from
the virial radius~\citep{opp10,vdvoort10}.  This effect becomes
stronger at low-$z$ as feedback energy accumulates in the IGM, as
evidenced e.g. by a much higher fraction of cosmic gas in the
Warm-Hot Intergalactic Medium in this model~\citep{dav10}.  At
$M_*=10^{10}M_\odot$, recycling is rare in this wind model~\citep{opp10},
so the net effect is to strongly lower the sSFR relative to the
no winds case.

In contrast, the momentum-driven wind scalings model increases the
sSFR relative to no winds.  In this case, the relationship to
no-winds is not so straightforward to understand, because the mass
loading factor depends on galaxy mass and epoch.  Overall, the
amount of energy input in vzw is smaller than in cw.  Recyling is
dominant at late times at $M_*\sim 10^{10}M_\odot$, but at early
epochs it is not infrequent.  The net effect is larger suppressions
of the sSFR at early epochs, while at late times the wind recycling
at this mass causes the sSFR to be higher relative to a simple
inflow model extrapolated from high-$z$.  Hence the vzw model yields
a shallower evolution of sSFR($z$) because it suppresses star
formation at early epochs and later recycles that material into
galaxies.

Observations show a very rapid rise from $z=0\rightarrow 2$, and
then essentially no evolution out to very high
redshifts~\citep{dad07,sta10,gon10}.  \citet{bou10} showed that
this is difficult to reproduce in a simple accretion-driven scenario,
and they argued for a rather radical modification to the galaxy
formation paradigm in which no stars formed in halos with virial
masses less than $10^{11}M_\odot$ (as well as above $10^{12}M_\odot$).
This strongly suppresses early star formation until the characteristic
mass of star-forming halos exceeds this threshold mass, at which
point it reverts to the accretion-driven scaling.  Clearly this
hypothesis is incorrect today; there is no evident feature in the
galaxy population around halo masses of $\sim 10^{11} M_\odot$,
unlike at $\sim 10^{12} M_\odot$ where many transitions occur.  It
is also difficult to accomodate at high-$z$, as smaller halos need
to form stars to reionise the Universe and form observed high-$z$
galaxy populations~\citep{mun11}.  Nevertheless, while implausible,
this scenario correctly emphasizes that some process must suppress
high-redshift star formation significantly in order to match current
sSFR evolution observations.

The observations are themselves uncertain to some extent, particularly
at high redshifts.  For instance, the original $z\sim 2$ data point
(black circle) was taken from dust-corrected UV-derived star formation
rates of star-forming BzK galaxies~\citep{dad07}.  Recent {\it
Herschel} data characterizing the infrared continuum of such sources
suggests that the extinction corrections were overestimated by up
to $\sim\times 2$~\citep{nor10}; this modification is shown by the
red arrow and circle in Figure~\ref{fig:ssfrevol}.  Prior to this
correction, the data exceeded the accretion rate evolution, making
it difficult to reconcile this evolution within the cold accretion
paradigm.  This prompted \citet{dav08} to make the radical suggestion
that the IMF may evolve with redshift, so that the true SFR's were
lower by $\sim\times 3$.  If the {\it Herschel}-derived correction
is right, the $z\sim 2$ data point is now consistent with being
powered by ongoing accretion, making it viable within the cold
accretion paradigm.  \citet{dav10b} pointed out that there still
may be a problem with respect to the momentum-driven scalings model;
this may be particularly true if the corrections at these masses
(which are not directly constrained by {\it Herschel} data) are
less than a factor of two~\citep[see also][]{nor10}.  But this is
no longer a fundamental difficulty, and this level of discrepancy
could in principle be alleviated by modifications of the feedback
prescription.  Indeed, \citet{genel11} find that doing high-resolution
disk simulations using a momentum-conserved feedback model with
somewhat different parameters results in galaxies that are in very
good agreement with (corrected) sSFR data at $z\approx 2$.

The slope $\beta$ of the main sequence (defined by SFR$\propto
M_*^{\beta}$) is also observed to evolve downwards from being unity
at high-$z$~\citep{sta10,lab10}, to $\beta\approx 0.9$ at $z\sim
1-2$~\citep{dad07,elb07}, to $\beta\approx 0.7$ at $z\la 1$~\citep{noe07}
down to $z\sim 0$~\citep{bri04}.  There is some sensitivity to
selection effects, particularly in how one chooses a sample of
star-forming galaxies at the massive end; the preponderance of
low-sSFR massive galaxies particularly at low-$z$ pulls the slope
down depending on how many such systems are included in the fit.  Hence
it is not straightforward to compare to our models where we do not
reproduce large passive galaxies.  In general, our momentum-conserved
winds case produces slopes of $\beta\approx 0.9-1$, roughly
independently of redshift, and the no-wind case produces a shallower
slope.  For the constant-$\eta$ models $\beta$ is ill-defined as
they show a strong three-tiered behavior.  The trend of the slope
becoming more shallow with time is therefore not obviously evident
even in our favored model.  While this may be a consequence of the lack
of quenching feedback in massive galaxies, we note that the
semi-analytic model of \citet{som08} includes AGN feedback but still
produces a slope around unity for the present-day main sequence.

In summary, the evolution of the specific star formation rate encodes
information about the feedback processes that have regulated star
formation up to that epoch.  Hence this is a critical quantity to
measure with upcoming surveys.  The fundamental driver is the gas
accretion rate onto halos, which evolves as $(1+z)^{2.25}$, but
feedback can modify this by either retarding accretion onto the
galaxy or adding recycled wind accretion.  Matching observations
that suggest a significant departure from this scaling at $z\ga 2$
requires suppressing star formation at early epochs.  Of the feedback
models examined here, only the momentum-conserved scalings case works in
this sense, though it is still well away from the unevolving sSFR
seen from $z\sim 4-7$.  Observations at these epochs remain fairly
uncertain~\citep[see e.g.][]{schaerer10}, so we await more robust
constraints from upcoming surveys.

\section{Satellite vs. Central Galaxies}\label{sec:sat}

The equilibrium paradigm for galaxy evolution introduced earlier
is appropriate for central galaxies, because cold accretion channels
material to the centers of halos.  Satellite galaxies, in contrast,
must scavenge their fuel from ambient halo gas, and therefore can
be impacted by a range of processes related to their surrounding
environment.  Satellite galaxies could thus display a broadly
different class of properties than central galaxies, for which
environment is not a major driver except through wind recycling.
Here we investigate the properties of satellite vs.  central galaxies
in our simulations, contrasting the statistics we have examined
earlier between the satellite and central galaxy populations.  We
particularly focus on dwarf galaxies where many environmental
processes could play a large role.

\subsection{Satellite Fraction}

\begin{figure*}
\vskip -0.2in
\setlength{\epsfxsize}{0.85\textwidth}
\centerline{\epsfbox{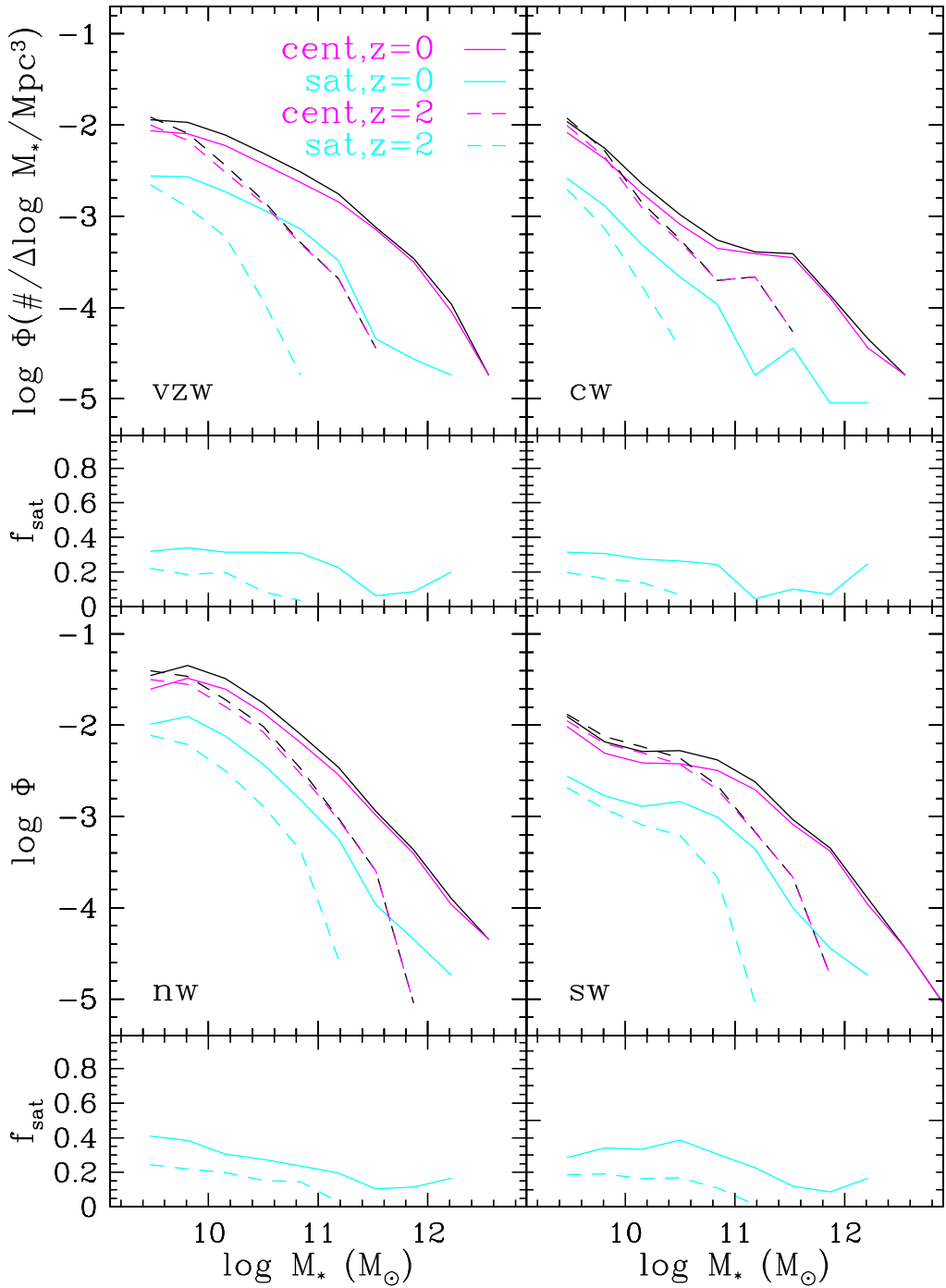}}
\vskip -0.5in
\caption{Four larger panels show stellar mass functions at $z=0$
(solid lines) and $z=2$ (dashed lines) for our four r48n384 simulations.
Black lines are the total mass functions (reproduced from
Figure~\ref{fig:massfcn}), which are subdivided into into centrals
(magenta) versus satellites (cyan).  Accompanying smaller panels below
each large panel show the fraction of satellite galaxies as a function 
of $M_*$ at $z=0$ (solid) and $z=2$ (dashed).
}
\label{fig:mfsat}
\end{figure*} 

Figure~\ref{fig:mfsat} shows stellar mass functions as in
Figure~\ref{fig:massfcn} (black curves), subdivided into central
galaxies (magenta curves) and satellite galaxies (cyan curves).
Solid curves show $z=0$ results, while dashed curves show $z=2$.
Smaller panel below each main panel show the fraction of satellite
galaxies as a function of stellar mass, also at $z=0$ (solid) and
$z=2$ (dashed).  Only the r48n384 series of runs are shown in this
plot.

Satellite galaxy mass functions are, for the most part, simply
scaled-down versions of the central galaxy mass functions.  They
display the same features arising from differential wind recycling.
In general, the properties of satellites appear to be established
when they are central galaxies in their own halo, and are not
dramatically affected by falling into a larger halo.  This is broadly
consistent with trends inferred from halo occupation distribution
models of galaxy clustering within dark matter halos~\citep[e.g.][]{con06}.

For all models at $z=0$, central galaxies dominate by number at all
resolved masses.  This may be surprising, as it is often assumed
that most small galaxies are satellites, and that the fraction of
satellites increases strongly to small masses.  The no-wind model
does show an increase in satellite fraction to lower masses, but
only mildly so.  In contrast, the wind models show a satellite
fraction that is approximately constant at one-third of all galaxies
for $M_*\la M^\star$.  For $M_*>M^\star$ the satellite fraction
decreases, although it is high again at very large masses -- these
are satellites in galaxy groups where the halo occupation distribution
is steeply rising~\citep[e.g.][]{ber03}.  These results agree with
those obtained by \citet[see their Figure~8]{vbosch08} based on the
conditional luminosity function constrained by the 2dF Galaxy
Redshift Survey, and recent SAMs also yield similar trends~\citep{guo10}.
At $z=2$, there are even fewer satellite galaxies at a given stellar
mass, typically only around 20\% of all galaxies for $M_*\la
10^{11}M_\odot$.  Hence dwarf galaxies are not typically satellites,
but rather central galaxies in small halos.

The relatively constant (and low) satellite fraction has a significant
implication for galaxy formation.  It means that that whatever
physics is responsible for establishing the mass function and other
such ensemble-averaged observables must predominantly govern central
galaxies, not satellites~\citep[see also][]{fon09}.  It is not
possible, for instance, to alter satellite galaxy physics to reconcile
the steep faint-end GSMF slope in the constant-$\eta$ or no-wind
simulations with observations.  This is even more true at higher
redshifts.  Hence the evolution of the ensemble population of
galaxies is governed by central galaxies, and understanding galaxy
evolution is, to first order, tantamount to understanding how central
galaxies form and grow.

\subsection{Quenched Satellites}\label{sec:satquench}

\begin{figure*}
\vskip -0.2in
\setlength{\epsfxsize}{0.85\textwidth}
\centerline{\epsfbox{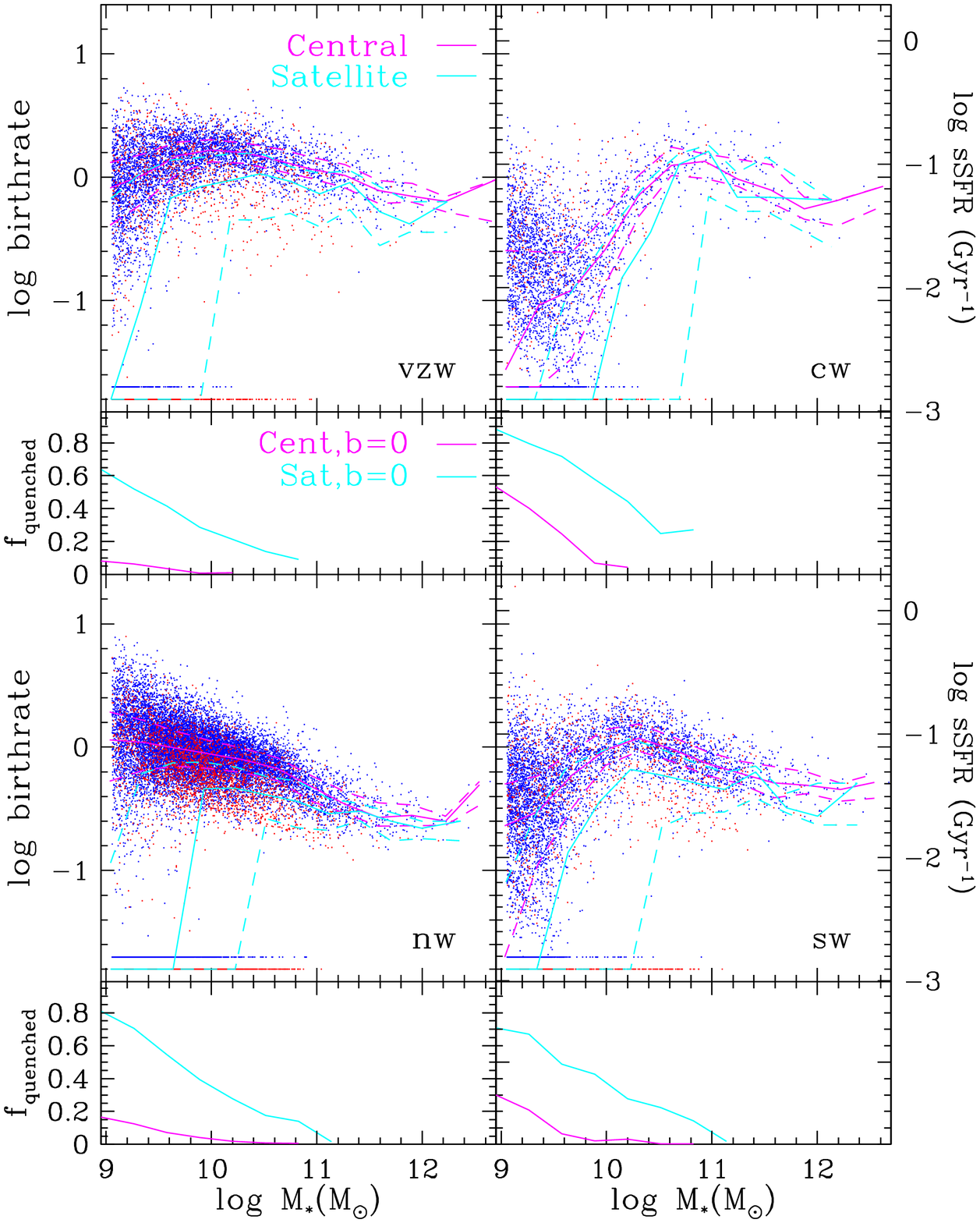}}
\vskip -0.5in
\caption{Larger panels show the birthrate (i.e. $t_{\rm SFR}$SFR/$M_*$,
where $t_{\rm SFR}=t_{\rm Hubble}-1$Gyr) for galaxies from
our r48n384 runs at $z=0$
color-coded by central (blue) and satellite (red) in our four
wind models.  Galaxies
with SFR$=0$ (``quenched") are shown along the bottom of the plot.
Running median (solid curves), including quenched galaxies, are
shown for central galaxies (magenta) and satellites (cyan).  Dashed
above and below show running 75\% and 25\% percentiles, respectively.
Small panels below show the fraction of quenched galaxies 
(having birthrate $b=0$) in a given mass
bin for centrals (magenta) and satellites (cyan).
}
\label{fig:ssfrsat}
\end{figure*} 

Figure~\ref{fig:ssfrsat} shows birthrates for galaxies in our four wind
models, separated into central (blue) and satellite (red) galaxies.
The birthrate is defined as the specific star formation rate multiplied
by the time over which the galaxy has been forming stars; the sSFR values
are noted on the right axis, for comparison with Figure~\ref{fig:ssfr}.
\citet{dav08} showed that the timescale for star formation in most
galaxies is reasonably well approximated by the Hubble time less 1~Gyr,
since none but the largest galaxies (today) form stars in the first Gyr.
A running median birthrate and lines enclosing 25\% and 75\% of galaxies
in a given mass bin are given by the solid and dashed lines, for centrals
(magenta) and satellites (cyan).

Birthrates are typically around unity for all models.  At the massive
end all models show a negative slope in birthrates with $M_*$,
indicative of archaeological downsizing.  At small masses, this can
invert owing to differential wind recycling as discussed in the
previous section.  Overall, however, star-forming galaxies show
generically smooth and fairly constant SFHs over much of cosmic
time to $z\sim 0$~\citep{dav08}, similar to that inferred for the
Milky Way.

There are a significant number of galaxies with no ongoing star formation
(i.e. ``quenched"), predominantly at low masses.  The smaller panels show
the fraction of quenched central (magenta) and satellite (cyan) galaxies
as a function of $M_*$.  There are a few small central galaxies with zero
instantaneous SFR's.  It is not clear what these are; one possibility
is that they are actually quenched satellites whose orbits carry them
outside the virial radius calculated by our spherical overdensity halo
finder, and therefore are identified as centrals.  Another possibility
is that quenching is partly numerical, since this may impact a satellite
galaxy's ability to cool gas from a hot halo.  Since they are very few
in number, we will leave a detailed investigation of these possibilities
for the future.

The satellite quenched fractions become large towards small masses.
Below a few$\times 10^9 M_\odot$, the median sSFR drops to zero for
satellites in all models.  Even above this mass, satellites show
lower typical sSFR's than central galaxies, hence environmental
processes can affect even large satellite galaxies.  The different
wind models (including no winds) show only minor differences in
terms of quenched satellite fraction, despite significant differences
in sSFR.  This indicates that it is not outflows that are responsible
for quenching satellites, and instead is some process(es) related
to environment that depends more on the growth of structure common
to all models.

Possible environmental processes that quench satellites are ram
pressure stripping, tidal harrassment, and strangulation (i.e.
cutting off of the gas supply).  \citet{sim09} showed that, in SPH
simulations similar to ours, quenching of a galaxy occurs on a $\sim
1$~Gyr timescale after entering another galaxy's halo.  The process
is gradual because subhalos retain their identity for quite some
time after entering a larger halo, so the satellite does not
immediately see the full effect of the hot gas in the larger halo.
In the cold accretion paradigm, the gas cooling along filaments
tends to fall to the center of the halo, and hence galaxies that
turn into satellites become disconnected from their feeding filaments.
But strangulation cannot be the entire story, because the gas
consumption timescales for these systems are of order a few Gyr (as
we will show in Paper~II), whereas quenching occurs on a timescale
of $\la 1$~Gyr.  Hence harrassment and stripping must be playing a
role.  Unfortunately, while the exact processes that drive satellite
quenching are interesting, they are a great numerical challenge to
model properly~\citep[e.g.][]{age07}, and the numerical resolution
and methodology used here is probably insufficient to make robust
quantitative statements.

Although there are seen to be a significant number of dwarf galaxies with
little or no ongoing star formation, SDSS data~\citep{wei06,vbosch08}
actually indicate a trend opposite to that predicted here, namely a
smaller fraction of red satellites (and centrals) at smaller masses.
One difference may be that these analyses defined red based on a color cut
between the red sequence and blue cloud, which does not straightforwardly
correspond to being fully quenched or not; the observed galaxies may
simply have lower specific SFRs.  Interestingly, a recent semi-analytic
model that is more successful than our simulations at matching a wider
range of data still suffers from the same discrepancy, showing an upturn
in red fraction at low masses~\citep{kim09}.  Hence the qualitative
discrepancy may hint at generic difficulties in modeling environmental
effects in satellites within hierarchical models, and bears further
investigation.

In summary, the fraction of satellite galaxies in our wind models
is roughly constant with mass for $M_*\la M^\star$ at around 20-30\%,
declining slightly with redshift.  Their mass functions show trends
similar to centrals, likely reflecting the fact that these trends
are set when satellites were centrals.  Satellites are increasingly
quenched to smaller masses, reflecting environmental processes that
gradually disconnect these galaxies from their feeding cold streams.
High-mass quenching must arise from another source that is currently
not included such as AGN feedback, but low-mass quenching can be
understood as a result of environmental effects that are already
included (at least in principle) in these types of models.

\subsection{Stellar Ages and Dwarf Galaxy Evolution}\label{sec:age}

\begin{figure*}
\vskip -0.2in
\setlength{\epsfxsize}{0.85\textwidth}
\centerline{\epsfbox{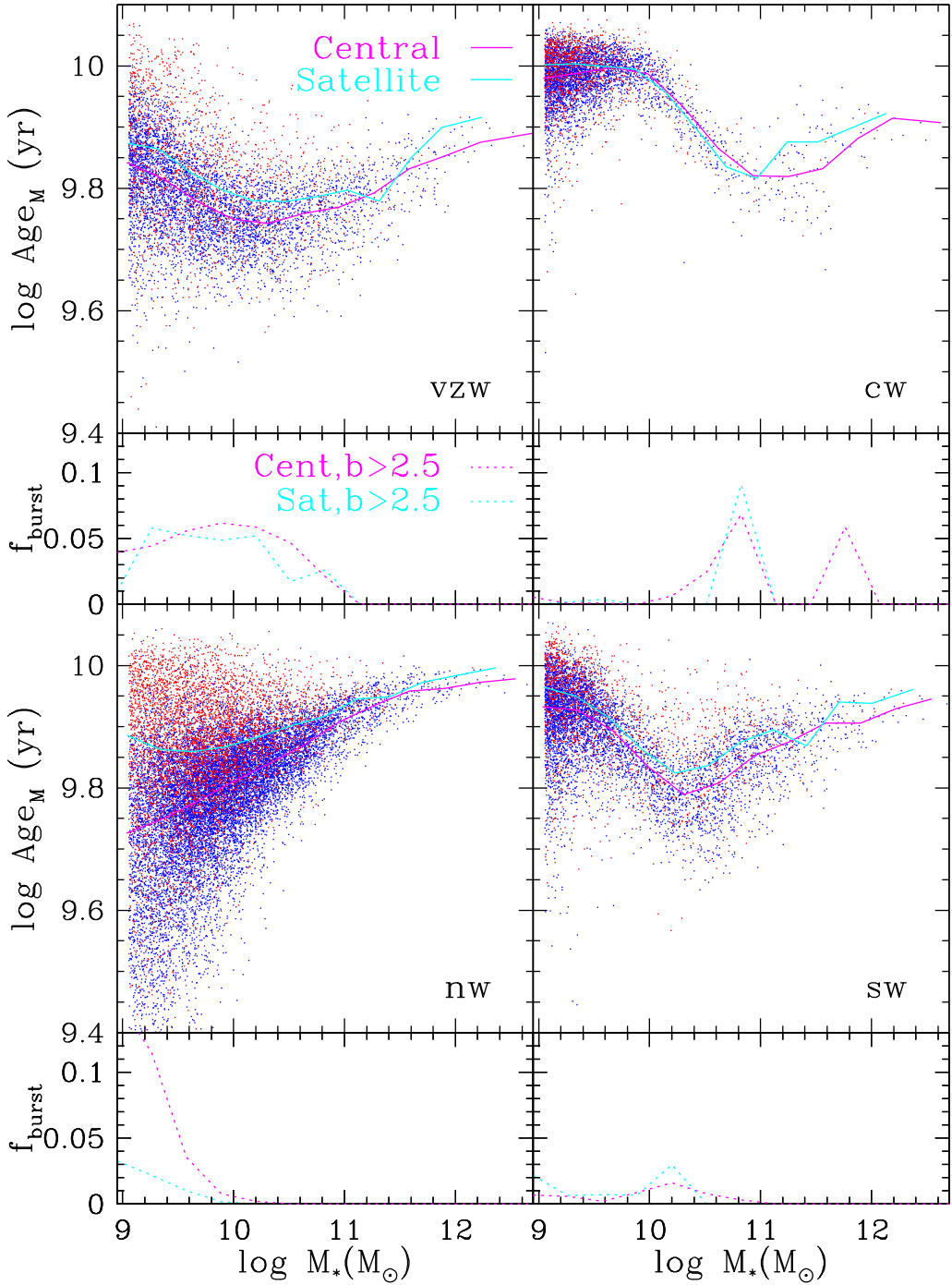}}
\vskip -0.5in
\caption{Larger panels show mean mass-weighted age of stars in galaxies
in our r48n384 runs
as a function of $M_*$, color-coded by central (blue) and satellite
(red) in our four wind models at $z=0$.  Running median (solid
curves), including quenched galaxies, are shown for central galaxies
(magenta) and satellites (cyan).  The smaller panels show the fraction
of satellites identified as ``bursting", i.e. with birthrate $b>2.5$,
as a function of mass, for centrals (magenta) and satellites (cyan).
}
\label{fig:agesat}
\end{figure*} 

We have seen that below masses of $M_*\la 10^{10} M_\odot$, the
wind models show a turn-down in sSFR and birthrate (and in Paper
II we will see a similar turn-down in gas fraction), for both
satellites and centrals.  This is in contradiction with
observations that show no such turn-down.

To investigate this further, we show in Figure~\ref{fig:agesat} the
mean (mass-weighted) stellar ages of galaxies as a function of
$M_*$, for satellites (red) and centrals (blue), in our four wind
models at $z=0$.  This is essentially an archaeological downsizing
plot, showing whether more massive galaxies have older stellar
populations as generally observed.  Running medians are shown for
centrals (magenta) and satellites (cyan).

All models exhibit archaeological downsizing for galaxies at the massive
end.  But for the wind models, below some mass the trend reverses,
and smaller galaxies begin to be progressively older.  This is opposite
to the trend observed by \citet{pas10} in a group catalog derived from
SDSS (see their Figure~4).  For the vzw and sw models, this happens at
$M_*\sim 10^{10} M_\odot$; above this mass, the ages are in reasonable
agreement with data.  They also show that satellites are slightly older
than centrals at a given mass, which is also qualitatively as observed,
although \citet{pas10} find the median mass-weighted age difference at
$M_*=10^{10} M_\odot$ is $\sim 0.1$~Gyr (becoming smaller to higher masses)
while our wind models predict a smaller difference.  The no-wind case 
shows an archaeological downsizing trend at all masses.

The mass where the turn-down happens is exactly the mass where the
specific star formation rate turns over (Figure~\ref{fig:ssfrsat}).
We argued previously that this owes to wind recycling, in particular
the lack of it in the smallest systems.  Hence we begin to build a
consistent story for the failure of wind models in the dwarf galaxy
regime: The smallest galaxies in these models begin forming stars
early, and owing to ejection of mass they lose their gas and cannot
reacquire it as fast as larger galaxies.  This is not because small
galaxies are satellites: central galaxies show exactly the same
trend, and most small galaxies are centrals anyway.

One possibility is that the simulations do not resolve the (supposedly)
bursty star formation histories of dwarf galaxies, which could give
rise to younger ages.  It is true that our simulations lack the
resolution to model internal resonances that can drive gas towards
the center of galaxies during interactions, fueling a burst~\citep{mih96}.
However, recent data has strongly questioned whether bursts are
common in dwarfs.  \citet{lee09} used the volume-limited 11HUGS
sample to show that the burst fraction, defined as having
birthrate$>2.5$, is only $6^{+4}_{-2}\%$ in dwarfs (typically $\la
0.1L^\star$), comprising $23^{+14}_{-9}\%$ of that population's
star formation.  These numbers are similar to what is seen for 
more massive galaxies~\citep[e.g.][]{bri04}

In the small panels in Figure~\ref{fig:agesat} we show the burst
($b>2.5$) fraction as a function of mass, for centrals (magenta)
and satellites (cyan).  It is generally quite low, and in our favored
vzw model it is mostly independent of mass at $\sim 5\%$ for $M_*\la
10^{10.5}M_\odot$, in both satellites and centrals.  In the no wind
case, the burst fraction rises dramatically at small masses, but
mostly in central galaxies not satellites.  The three-tiered shapes
of the birthrate curves in the cw and sw model result in odd behaviors
in the burst fraction, and result in very low burst fractions for
small systems.  The favoured momentum-conserved wind case agrees
well with observations of burst fractions, although the fraction
of star formation in bursts is $\sim 15$\% which is $\approx 1\sigma$
below that observed.  This model's concordance with data suggests
that, contrary to popular wisdom, small dwarfs are not significantly
burstier compared to larger star-forming systems.

The higher ages, steep high-$z$ GSMF, and lower sSFR's for dwarfs
suggest that wind models should suppress early star formation in
smaller systems even more than is currently done in the momentum-conserved
wind case, and produce a larger reservoir of gas for consumption
at later epochs.  This could arise if mass loading factors are
larger at early epochs than assumed here, and/or wind speeds are
reduced at late times such that these galaxies can reacquire their
ejected gas.  This could also arise if the star formation law is
different in these systems owing to a less efficient conversion of
atomic gas into molecular~\citep[e.g.][]{rob08} -- this would tend
to delay star formation, providing a younger, more vigorously
star-forming dwarf population today.  Investigating this discrepancy
further is likely to reveal new physical processes that are important
for the evolution of small galaxies at all cosmic epochs.

\section{Summary and Discussion}\label{sec:summary}

Cosmological hydrodynamic simulations are now reaching a maturity
level such that they can be informatively compared to observations
of galaxies across cosmic time.  This is a new era in this type of
modeling, as past such efforts have generally produced quite poor
agreement with data, and have been severely limited by dynamic range
particularly when evolved to low redshifts.  With improvement in
both computing power and input physics, simulations can now plausibly
reproduce a wider suite of observed relations (though still far
from all), which in return can yield insights into the governing
physics.  A particularly important new physical aspect has been the
incoporation of galactic outflows, originally motivated to explain
IGM enrichment, and simultaneously having a wide-ranging impact on
galaxies' stellar, gas, and metal content. 

In this paper we present a study of how the stellar content and
star formation rates of galaxies are impacted by galactic outflows.
We concentrate on understanding the underlying physical drivers of these
properties, and how observations can enlighten us on the way in which
these drivers operate in the real Universe.  To this end we present
some comparisons with observations, showing that including outflows
with momentum-conserved scalings provide the best overall match to
the ensemble of observations considered in this work, although notable
discrepancies remain.  Given the broader success of this class of models
in matching IGM (and other galaxy) properties, and that these scalings
are consistent with those observed for outflows, this adds to the growing
body of evidence that strong and ubiquitous outflows with these scalings
are an important piece in understanding the overall evolution of galaxies.

The evolution of the stellar component can be broadly understood within
the context of a cycle of gas inflow and outflow between galaxies and
the IGM.  This differs somewhat from the traditional view of galaxy
formation in which halos and their mergers drive galaxy evolution.
The cycle of baryons can at its most basic level be analytically
represented by a slowly-evolving equilibrium between inflow, outflow,
and star formation (eq.~\ref{eqn:equil}), where the outflows govern
how much of the inflowing material turns into stars.  The galaxy
mass dependence of the outflow rate thus directly governs e.g. the
GSMF, the specific star formation rate of galaxies, and galaxy ages.
A particularly key aspect is the idea of differential wind recycling, in
which material ejected returns to galaxies in a mass-dependent manner.
This drives noticeable features in observable properties of galaxies
as a function of mass, providing a way to constrain how outflows move
material within the cosmic ecosystem surrounding galaxies.  We reiterate
that this scenario does not invoke the halo virial radius, environment,
or mergers as playing a governing role in star-forming galaxy evolution.
The cold streams feeding star-forming galaxies generally take no notice of
the virial radius, are broadly unaffected by environment, and while such
streams also carry in galaxies that merge, the mass growth rate in such
mergers is sub-dominant~\citep{dek09}.  In contrast, the virial radius,
environment, and mergers seem to play a pivotal role in quenching star
formation, and more generally in the evolution of massive halos hosting
quiescent galaxies.

With that framework in mind, we summarize the key conclusions of this
paper:
\begin{itemize}

\item Galactic outflows are required to suppress stellar mass growth at
all epochs.  The stellar mass function is quite steep and nearly a power
law at high redshifts. At lower redshifts it develops a more pronounced
three-tier behavior, where the middle tier reflects the steepness of
differential (i.e. mass-dependent) wind recycling~\citep{opp10}, and the
higher and lower tiers tend towards the slope of the halo mass function.
The middle tier (typically just below $M^\star$) becomes shallower with
time owing primarily to wind recycling, which returns ejected material
faster to more massive galaxies.  Observations of the $z\approx 0$
GSMF also show a three-tier behavior, and the momentum-conserved wind
simulation comes closest to matching its behavior (although the three
tiers are not obviously evident).

\item The star formation rate functions show many of the same general
features as the GSMF, but the detailed evolution, mass dependence,
and dependence on outflows are somewhat different.  Comparisons of
these models to present-day data are hampered by their lack of
quenching feedback in massive galaxies in our models.  At $z\ga 1$
where massive quenched galaxies are less common, momentum-conserved
winds produces a good match to the observed star formation rate
function derived from H$\alpha$ data.

\item The specific star formation rate's dependence on mass is
governed by how the effective mass loading factor of outflows varies
with stellar mass.  The no-wind case shows too little star formation
at a given $M_*$, despite having far too much star formation globally.
The momentum-conserved wind model matches observations well at
$\sim 0.1-1\;M^\star$, but all wind models show too little star
formation at the smallest masses.  This may indicate that there
should be less mass loading or more wind recycling at the $M_*\la
0.1 M^\star$ compared to what is currently in these models, or
else the conversion of gas into stars is not being modeled properly
in low-mass systems.

\item The evolution of the sSFR is generally well-described as
being driven by cosmic mass accretion into halos, which scales as
$(1+z)^{2.25}$.  The no-wind and slow-wind cases follow this trend at
all redshifts.  Observations do so at $z\la 2$, but at $z\ga 2$ they
depart from this scaling, indicating that small high-redshift galaxies
must be significantly suppressed.  The momentum-conserved wind model
qualitatively yields this, but not enough to match current high-$z$
observations.  Pinning down the sSFR evolution at $z\ga 2$ will have a
major impact on understanding early galaxy evolution and feedback
processes.

\item Satellite galaxies are the minority of galaxies at all masses
probed by these simulations (down to $M_*\sim 10^9M_\odot$), with
a fraction of around one-third at $z=0$ independent of $M_*$.  Hence
understanding the growth of galaxies must focus on developing a
model for central galaxies.  The satellites' GSMF follows the same
trends as that of centrals.  However, there are many more non-star
forming satellites than centrals, particularly at lower masses,
owing to environmental effects wherein the satellites no longer
effectively receive inflow from the IGM; this trend may be contrary
to that observed.  Small galaxies are at most only mildly more
bursty than larger ones, and the burst fraction is small in accord
with observations.  Our results challenge conventional notions that
most small galaxies are bursty satellites.

\end{itemize}

Taking a broader view, it has long been recognised that the galaxy
formation process must provide some feedback mechanism that preferentially
suppresses mass growth at both high and low masses.  In this paper
we have advocated that it is galactic outflows that are primarily
responsible for regulating the low-mass end.  The advantages of invoking
this particular mechanism are that (i) outflows are observed and seem
to follow scalings that yield many desirable properties in models; (ii)
they concurrently enrich the IGM at all cosmic epochs as observed; (iii)
they lower the baryon content of halos preferentially to lower masses as
observed~\citep[e.g.][]{dav10d}; and (iv) they are an ejective feedback
mechanism that works in conjunction with the cold accretion paradigm,
in which it is difficult to prevent filaments from channeling gas
efficiently into galaxies.  There are other ideas to suppress low-mass
star formation by regulating it as a function of mass internally within
the ISM (i.e. by varying the star formation efficiency as a function
of $M_*$), or by preventing material entering the halo from accreting
into small galaxies.  But these scenarios must explain IGM enrichment
separately, and must hide most of the gravitationally-accreted baryons
in halos in some undetected form which is becoming increasingly difficult
to accomodate~\citep[e.g.][]{mcg10,dai10}.

While attractive, the idea of invoking outflows is not without
its difficulties.  For instance, the amount of mass being driven
out of these galaxies is large, globally many times the amount of
mass forming into stars.  This is a generic result, because some
process must suppress stellar masses by such large factors at the
low mass end~\citep{ker09a,opp10}; in our models, this process is
galactic outflows, as opposed to ISM physics or preventive feedback.
The physical mechanism(s) that would drive such a sizeable amount of
material out of galaxies is unclear.  The canonical idea of driving
outflows from overpressurized bubbles in the ISM tends to (when
modeled) create holes by which energy and metals, but little mass,
are ejected~\citep[e.g.][]{maclow99}, and there are difficulties
with entraining cold clouds as observed without disrupting them.
The scalings preferred by our current comparisons to data are those
expected for radiation-driven winds, but it is not well known how far out
the photons' momentum can drive dust, and how far out the dust remains
coupled to the gas.  For instance, \citet{yos10} use spectropolarimetry
to determine that the dust in M82 is moving outwards much more slowly
than the gas, suggesting that the gas and dust are not strongly coupled.
Furthermore, the momentum budget of typical stellar populations is, under
the assumption of single photon scattering, insufficent to drive outflows
as appear to be necessary, indicating that supernovae or other sources of
energy are still required.  Without a solid understanding of the dynamics
driving outflows, simulations such as the ones presented here cannot be
considered a complete description of the physics of galaxy formation.
Fortunately, observations of outflows across cosmic time are gaining
substantial traction~\citep{mar05,wei09,ste10}, providing empirical
constraints for models while the theory of outflow propagation develops.

Our momentum-conserved winds provides the best match to data
on star-forming galaxies of the four models considered here.
But the agreement is only good in the rather narrow mass range from
$10^{10}\la M_*\la 10^{11} M_\odot$.  At larger masses, the current
models require some additional quenching mechanism perhaps associated
with AGN~\citep[e.g.][]{dim05,cro06,gab11}.  At smaller masses,
this model produces too vigorous early star formation in dwarfs,
resulting in too-old stellar populations, too steep mass functions,
and insufficient star formation today~\citep[a result also seen in
many SAMs, e.g.][]{fon09,guo10}.  This may indicate that feedback in
these dwarf systems evolves in some way not captured by this particular
outflow prescription, or else that there is some additional physical
process that is important.  It is possible that the star formation
law is different in smaller systems to make gas consumption less
efficient~\citep[e.g.][]{rob08}, or else that these simulations have
too low a density threshold for star formation (owing to resolution
limitations) such that they do not properly represent star formation
in low surface brightness, gas-rich systems~\citep[e.g.][]{gov10}.
More broadly, none of our simulations can simultaneously reproduce
the shallow stellar mass function, archaeological downsizing, and
star formation rate downsizing as observed; this remains a fundamental
challenge for galaxy formation models.  Even though the concordant mass
range of the momentum-conserved wind model is fairly narrow, it is still
a significant step forward as it provides physical insights into the
processes that govern galaxy evolution at all masses, pinpointing both
successes and failures.  And it is worth recognizing that this particular
dex in stellar mass happens to be where most of cosmic star formation
occurs, so it is a critical mass range for understanding global stellar
mass growth.

Overall, the general simulation-inspired framework for how inflows
and outflows govern galaxy formation is compelling from both an
observational and theoretical perspective, and is well-situated
within the current hierarchical structure formation paradigm.  This
makes it an attractive scenario to pursue both analytically and
numerically.  In Paper~II we will examine how this framework provides
intuition about the gaseous and metal content of galaxies, thereby
together covering the primary constituents that determine the
observable multi-wavelength properties of galaxies at all cosmic
epochs.

 \section*{Acknowledgements}
The authors acknowledge A. Dekel, M. Fardal, M. Haas, N. Katz, D. Kere\v{s},
J. Kollmeier, C. Papovich, and D. Weinberg for helpful discussions,
S. Weinmann for useful comments on an early draft, J. Schaye for helpful
refereeing, and V. Springel for making \gad\ publicly available.
RD thanks North West University in Mafikeng, South Africa and the
University of Cape Town, South Africa for their hospitality during
much of the writing of this paper.  The simulations used here were run
on University of Arizona's SGI cluster, ice.  Support for this work
was provided by NASA through grant number HST-AR-11751 from the Space
Telescope Science Institute, which is operated by AURA,Inc. under NASA
contract NAS5-26555.  This work was also supported by the National
Science Foundation under grant numbers AST-0847667 and AST-0907998.
Computing resources were obtained through grant number DMS-0619881 from
the National Science Foundation.

\end{document}